\documentclass[preprint,sort&compress,5p]{elsarticle}
\usepackage{amsmath, amssymb, graphicx}
\usepackage[switch]{lineno}
\usepackage{hyperref}
\usepackage[hypcap]{caption}
\usepackage{subfig}
\usepackage{xcolor}
\hypersetup{colorlinks}
\usepackage[english]{babel}

\begin{document}

\title{The Detector Calibration System for the CUORE cryogenic bolometer array}
\author[Yale]{Jeremy~S.~Cushman\corref{cor}}
\ead{jeremy.cushman@yale.edu}
\author[Wisc]{Adam~Dally}
\author[Yale]{Christopher~J.~Davis}
\author[Wisc]{Larissa~Ejzak\fnref{fn1}}
\author[Wisc]{Daniel~Lenz}
\author[Yale]{Kyungeun~E.~Lim}
\author[Yale]{Karsten~M.~Heeger}
\ead{karsten.heeger@yale.edu}
\author[Yale]{Reina~H.~Maruyama}
\author[Milano,INFNMiB]{Angelo~Nucciotti}
\author[Wisc]{Samuele~Sangiorgio\fnref{fn2}}
\author[Yale,Wisc]{Thomas~Wise}

\cortext[cor]{Corresponding author}
\fntext[fn1]{Present address: Research Square, Durham, NC 27701, USA}
\fntext[fn2]{Present address: Lawrence Livermore National Laboratory, Livermore, CA 94550, USA}

\address[Yale]{Wright Laboratory, Department of Physics, Yale University, New Haven, CT 06520, USA}
\address[Wisc]{Department of Physics, University of Wisconsin, Madison, WI 53706, USA}
\address[Milano]{Dipartimento di Fisica, Universit\`{a} di Milano-Bicocca, Milano I-20126, Italy}
\address[INFNMiB]{INFN -- Sezione di Milano Bicocca, Milano I-20126, Italy}

\journal{Nuclear Instruments and Methods in Physics Research Section A}

\begin{abstract}
The Cryogenic Underground Observatory for Rare Events (CUORE) is a ton-scale cryogenic experiment designed to search for neutrinoless double-beta decay of $^{130}$Te and other rare events. The CUORE detector consists of 988 TeO$_2$ bolometers operated underground at 10~mK in a dilution refrigerator at the Laboratori Nazionali del Gran Sasso. Candidate events are identified through a precise measurement of their energy. The absolute energy response of the detectors is established by the regular calibration of each individual bolometer using gamma sources. The close-packed configuration of the CUORE bolometer array combined with the extensive shielding surrounding the detectors requires the placement of calibration sources within the array itself. The CUORE Detector Calibration System is designed to insert radioactive sources into and remove them from the cryostat while respecting the stringent heat load, radiopurity, and operational requirements of the experiment. This paper describes the design, commissioning, and performance of this novel source calibration deployment system for ultra-low-temperature environments. 
\end{abstract}

\maketitle

\tableofcontents

\section{Introduction}
The Cryogenic Underground Observatory for Rare Events (CUORE) is an experiment designed to search for neutrinoless double-beta ($0\nu\beta\beta$) decay of $^{130}$Te, with the goal of investigating the nature and mass of the neutrino~\cite{Arnaboldi:2004dd}. CUORE consists of 988 TeO$_2$ crystals~\cite{Arnaboldi:2010cw}, each with dimensions of $5\times 5\times 5\ \mathrm{cm^3}$, operated as bolometers inside a large, custom-designed cryostat~\cite{Nucciotti:2008be}. The bolometers are both the sources and detectors of $^{130}$Te decay. They are arranged in 19 towers --- $2\times2\times13$ arrays of crystals mounted in copper frames --- and are cooled to approximately 10~mK by a dilution refrigerator.

As a particle passes through a CUORE crystal, the energy it deposits is converted into phonons, causing a temperature rise that is measured by a neutron-transmutation-doped germanium thermistor~\cite{Haller:1984dr}. The relationship between the thermistor voltage reading and the original particle energy is nonlinear and unique to each bolometer--thermistor pair~\cite{Vignati:2010gp}. Because the signature of $0\nu\beta\beta$ decay is a peak in the energy spectrum at the $Q$-value of the decay, a precise understanding of the bolometer energy scale is critical for detecting this process. In addition, detecting other rare processes, such as two-neutrino double-beta decay, requires an understanding of the spectrum over a wide range of energies. As a result, absolute energy calibration of each bolometer using sources at a variety of energies is required. The response of the bolometers and thermistors is highly temperature-dependent, and as such, the calibration must be performed with the bolometers at their base temperature.

\sloppy 
During the projected 5-year operating period of CUORE, we will calibrate the bolometer--thermistor pairs regularly, as detector conditions can change over time. In \mbox{CUORE-0}, a predecessor experiment, calibration was performed monthly~\cite{Alfonso:2015vk}, and we expect the calibration frequency to be similar for CUORE. Each calibration period is kept as short as possible to maximize the live time for physics data taking. At the same time, because of the $\sim$4-second response and recovery time of the bolometers~\cite{Artusa:2014fn}, the source activity must be sufficiently low to avoid pile-up. Because of the compact configuration of the CUORE detector towers, the outer bolometer towers in the cryostat partially shield the innermost towers from external radiation. As a result, to achieve calibration periods of one or two days while not saturating the outer detectors, calibration sources must be placed near each bolometer throughout the tower array during calibration. Because these sources will be removed from the cold detector region of the cryostat during physics data taking, they must be cooled to the cryostat's base temperature for each calibration and subsequently warmed up again.

\fussy
Deploying calibration sources into a cold cryostat without significant disruption to the cryostat operating temperature poses demanding technical challenges. To accomplish this task, we have designed and implemented the CUORE Detector Calibration System (DCS). In \mbox{\autoref{sec:system_overview}}, we present an overview of the DCS, including the design and production of the calibration sources, the motion control and monitoring hardware, and the tubes and other hardware that guide the calibration sources through the cryostat. In \mbox{\autoref{sec:control_electronics}}, we discuss the electronic control system for the DCS and the remote and automatic DCS software controls. Finally, we present and discuss the results of a calibration source deployment down to base temperature in the CUORE cryostat in \mbox{\autoref{sec:dcs_performance}}.

\section{System overview}
\label{sec:system_overview}
\begin{figure}
\begin{center}\includegraphics[width=3.45in]{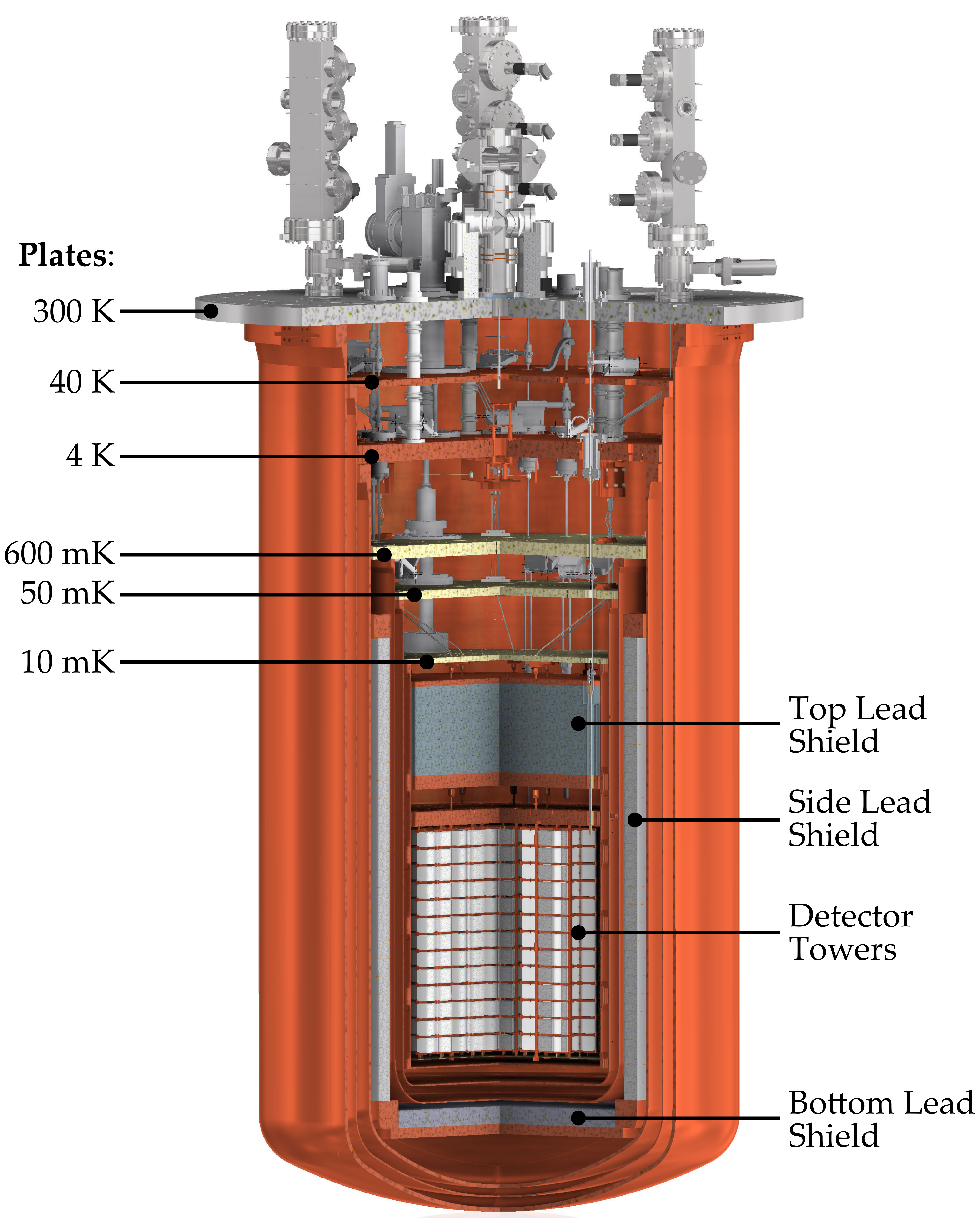}\end{center}
\caption{Illustration of the CUORE cryostat with a quarter cutout.}
\label{fig:cryostat}
\end{figure}

The CUORE cryostat contains a large custom-built cryogen-free dilution refrigerator assisted by pulse tube cryocoolers~\cite{Nucciotti:2008be}. It comprises six plates and corresponding copper vessels held at successively colder temperatures (see \mbox{\autoref{fig:cryostat}}). A stainless steel room-temperature (300~K) plate provides all connections to the outside of the cryostat and supports the calibration system and other hardware. Copper plates at 40~K and 4~K are cooled by pulse tube cryocoolers. A gold-plated copper plate at 600~mK is thermally coupled to the still of the dilution unit, and there are similar plates at 50~mK and at 10~mK, coupled to the heat exchanger and mixing chamber, respectively. The detector towers are located below the mixing chamber, underneath the top lead shielding. The DCS is the motion and thermalization hardware that guides calibration sources into the cryostat and extracts them after each calibration period has concluded.

\begin{figure}
\begin{center}\includegraphics[width=3.45in]{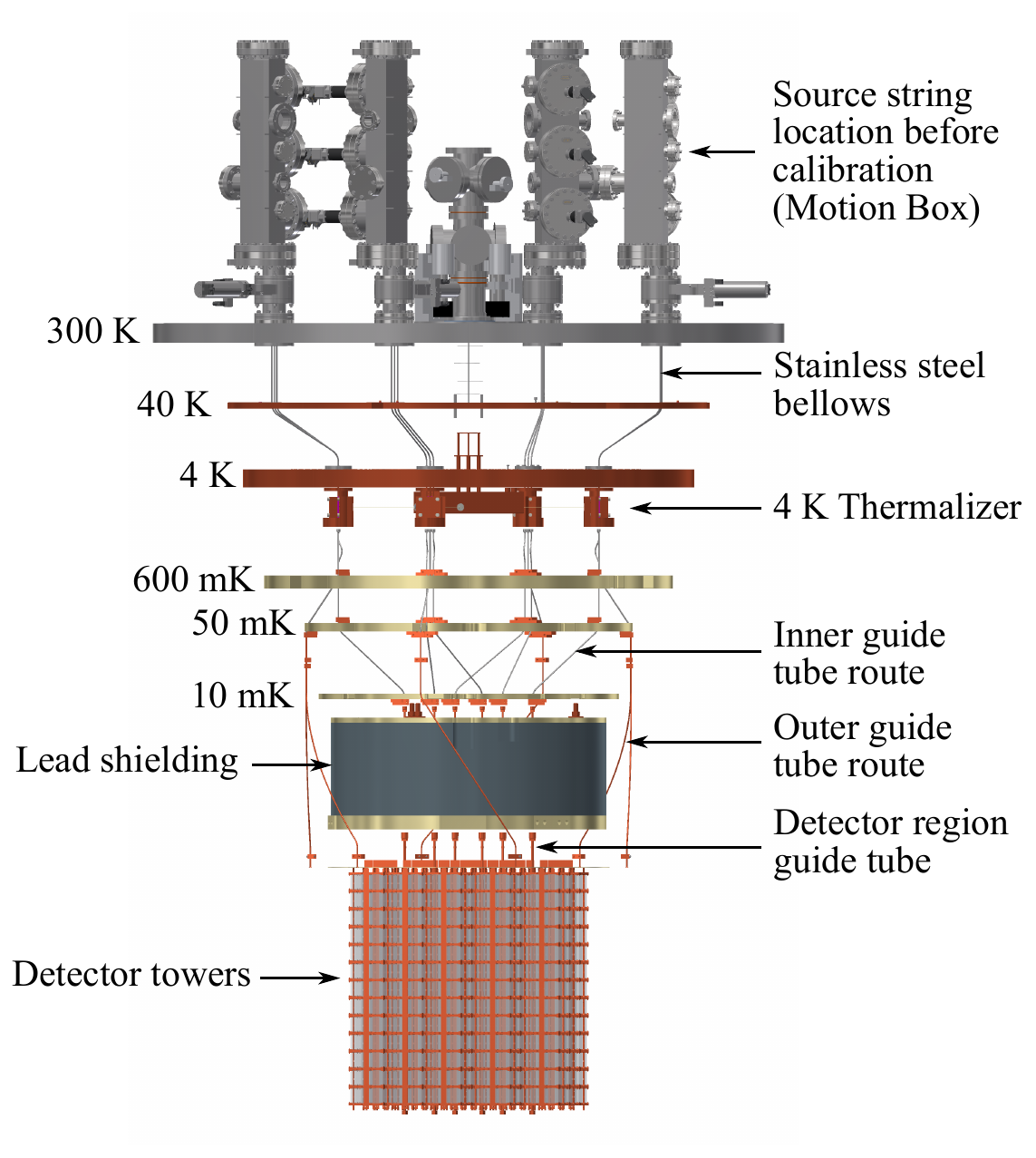}\end{center}
\caption{Illustration of the DCS in the CUORE cryostat.}
\label{fig:dcs_integration}
\end{figure}

Each calibration source carrier is a collection of individual source capsules attached to a continuous string. There are 12 calibration source carriers (``source strings'') in total. During physics data taking, these source strings are wound on spools above the cryostat at room temperature, outside of the internal lead and copper cryostat shielding, to avoid gamma rays from the strings reaching the detectors. The spools are contained inside vacuum-tight stainless steel enclosures that are connected to the inner cryostat vacuum through gate valves. At the beginning of each calibration period, motorized and computer-controlled spools lower the source strings under their own weight into the cryostat.

As the strings are lowered, they are guided through the cryostat by a series of polytetrafluoroethylene (PTFE), stainless steel, and copper tubes (collectively referred to as ``guide tubes''). As the strings pass the \mbox{4-K} stage of the cryostat, a thermalization mechanism consisting of two parallel, spring-loaded copper blocks squeezes the sources to cool them down to 4~K. Below this stage, the guide tubes divide into inner and outer  paths. The six inner paths bring strings to the innermost region of the cryostat where the detectors are mounted, whereas the six outer paths bring strings to just outside the \mbox{50-mK} vessel. The inner strings pass through lead shielding above the detectors and reach their final positions inside copper tubes mounted between the detectors; the outer strings are allowed to hang freely because they are outside the detector region. An illustration of the DCS in the CUORE cryostat is shown in \mbox{\autoref{fig:dcs_integration}}.

\subsection{Requirements}
The primary function of the DCS is to deploy calibration sources into the cryostat and cool them down without affecting the operating temperature of the detectors. The design and construction of the system is largely driven by the strict thermal and radioactivity requirements of the experiment. The DCS must respect the thermal load requirements of each stage of the dilution refrigerator, both when sources are stationary and when they are moving. The hardware that remains in the cryostat during physics data taking must also make a negligible contribution to the radioactive background in the $0\nu\beta\beta$-decay energy region of interest around 2528~keV~\cite{Redshaw:2009bg,Scielzo:2009co,Rahaman:2011wt}. Finally, the system must operate safely and stably over the lifetime of the experiment and must be flexible, allowing us to change or replace any of the calibration sources as necessary.

The source strings and guide tubes must only minimally impact the thermal conductivity between the various thermal stages of the cryostat. In addition, there must be no straight-line access between the very different temperature regions of the cryostat to minimize thermal radiation from warmer to colder stages. During string motion, frictional heating, which can be particularly problematic at the coldest stages of the cryostat, must also be minimized. Moreover, during string lowering, the source strings must be cooled as they are lowered to avoid dissipating large amounts of heat in the colder parts of the cryostat. Specifically, the DCS is designed such that any heat load due to source deployment and extraction can be compensated for by the temperature stabilization system of the cryostat, thereby avoiding any effects on the cryostat base temperature. The design goal for the cooling power available to the DCS in the cryostat is shown in \mbox{\autoref{tab:cooling_power}}.

\begin{table}
\begin{center}\begin{tabular}{c|c}Thermal stage & Available cooling power\\\hline
40~K & 1~W \\
4~K & 300~mW \\
600~mK & 600~$\mu$W \\
50~mK & 1~$\mu$W \\
10~mK & 1~$\mu$W
\end{tabular}\end{center}
\caption{Cooling power budgeted for the calibration system at all thermal stages of the cryostat.}
\label{tab:cooling_power}
\end{table}

Another set of requirements for the DCS arises from the low-background environment that is necessary for a rare-event search with CUORE. Most importantly, the detectors must be shielded from the radioactivity of the calibration sources during physics data taking, which necessitates removing the sources from the detector region. In addition, the radioactivity of the materials used to construct the DCS must be low, especially for the source string guide tubes in the detector region. To achieve the CUORE background goal of $10^{-2}~\mathrm{counts/(keV\,kg\,yr)}$ in the $0\nu\beta\beta$ region of interest, this necessitates constructing all DCS hardware from only ultrapure copper, with $^{232}$Th and $^{238}$U bulk contamination levels at or below $10^{-12}$~g/g, in the detector region of the cryostat; in this region, we construct the DCS hardware from the same copper used in the frames of the CUORE detector towers.

The DCS must also be both fail-safe and very unlikely to fail. The system must be constructed such that the risk of active source material escaping into or remaining behind in the cryostat is essentially zero. It must also be designed such that the source capsules do not become stuck in the cryostat. If any source material were to become stuck or remain behind in the cryostat following a calibration period, it would be necessary to warm up the cryostat to room temperature and open it to extract this material, which is a lengthy process. Therefore, we use a variety of sensors to monitor the system to ensure that any abnormal behavior will be caught before any damage occurs to the source strings or the cryostat.

Finally, the calibration sources must be replaceable without warming up the cryostat to allow a variety of different source isotopes to be inserted if desired.

\subsection{Calibration sources}
\label{sec:calibration_sources}
The $Q$-value of $^{130}$Te $0\nu\beta\beta$ decay is 2528~keV; thus, a variety of gamma decay lines are appropriate for calibrating all energies up to this $Q$-value. For CUORE, $^{232}$Th sources are a natural choice because of the wide variety of lines provided by the daughter isotopes of $^{232}$Th, including the strong $^{208}$Tl line at 2615~keV, which enables precise energy calibration near the $Q$-value. Sources containing $^{232}$Th were also used in Cuoricino and \mbox{CUORE-0}, predecessor experiments to CUORE, thus providing us with a vetted analysis framework for performing this calibration~\cite{Ejzak:2013tm, Alduino:2016ik}. The long half-life of $^{232}$Th, $1.4\times 10^{10}$ years~\cite{Browne:2006cg}, means that we will not need to replace the $^{232}$Th sources over the 5-year lifetime of the experiment.

The source material is commercially available thoriated tungsten wire that fits into small source capsules. The source activity is tuned by choosing the desired thorium composition of the wire (1\% or 2\% thorium by weight, nominal) and cutting it to a precise length. The source capsules accept wires of up to 5.1~mm in length and 0.50~mm in diameter. The activity of the uncut source wires was measured by the Berkeley Low Background Facility~\cite{Thomas:2013ia} to a precision of approximately 3\%.

\begin{figure}
\begin{center}\includegraphics[width=3in]{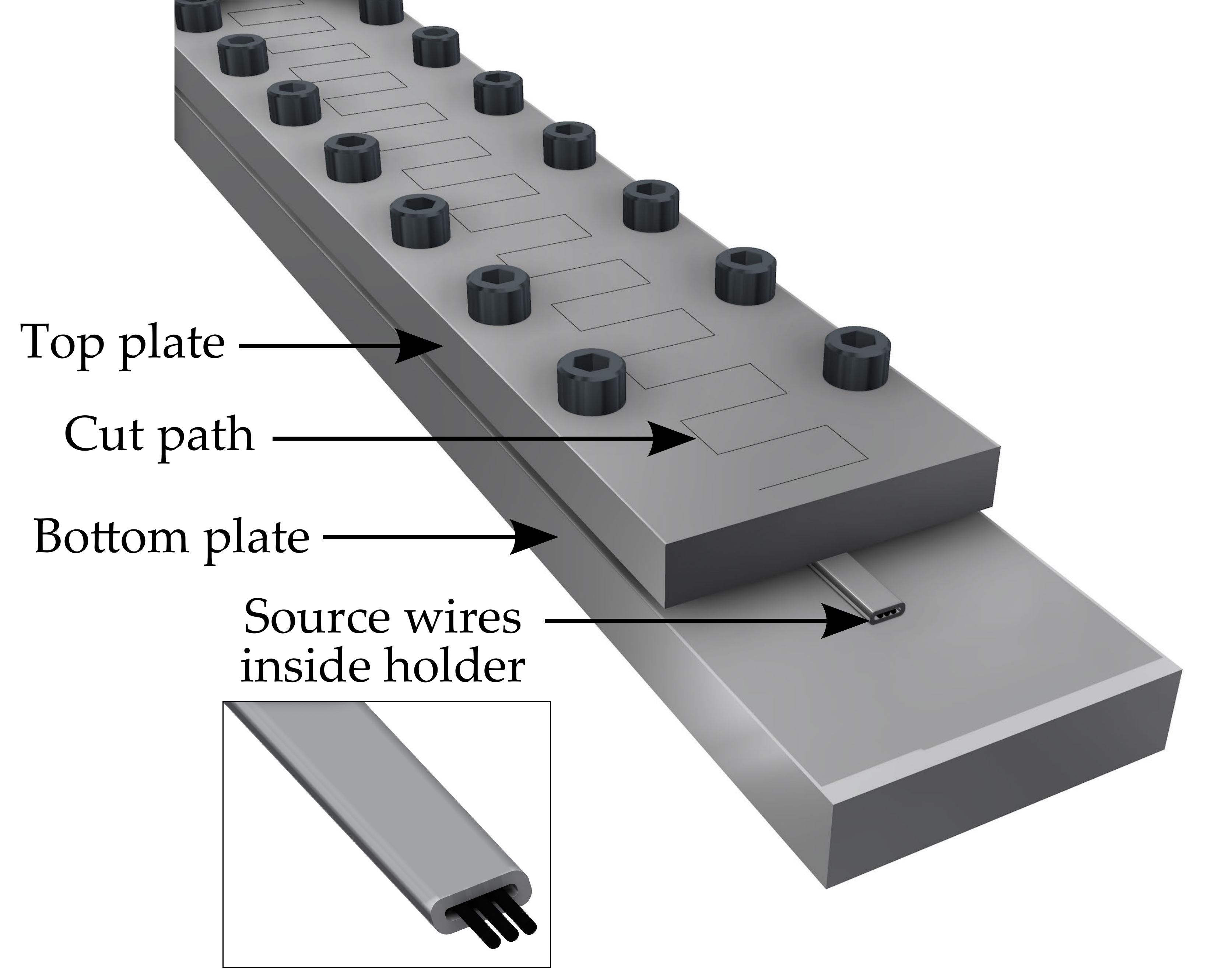}\end{center}
\caption{Rendering of the jig used for the wire electrical discharge machining of the source wires. The top and bottom plates and the source wire holder are aluminum.}
\label{fig:edm_jig}
\end{figure}

Wire electrical discharge machining makes it possible to cut the source wires reliably, with clean, hard corners and no fraying. Prior to machining, we place the source wires inside small aluminum holders, which are themselves placed inside a larger aluminum jig (see \mbox{\autoref{fig:edm_jig}}). Perpendicular cuts are made through the entire jig, with the sources inside, to produce wires of the desired length. The aluminum holders ensure that the source wires remain in place before and during this cutting. We then dissolve the aluminum holders in a strong sodium hydroxide (NaOH) solution and clean the source wires with citric acid (C$_6$H$_8$O$_7$). The cutting procedure has a precision of a few hundredths of a millimeter, corresponding to an additional uncertainty on the string activity of approximately 1\%.

\begin{figure}
\begin{center}
\subfloat[][]{\includegraphics[height=2.2in]{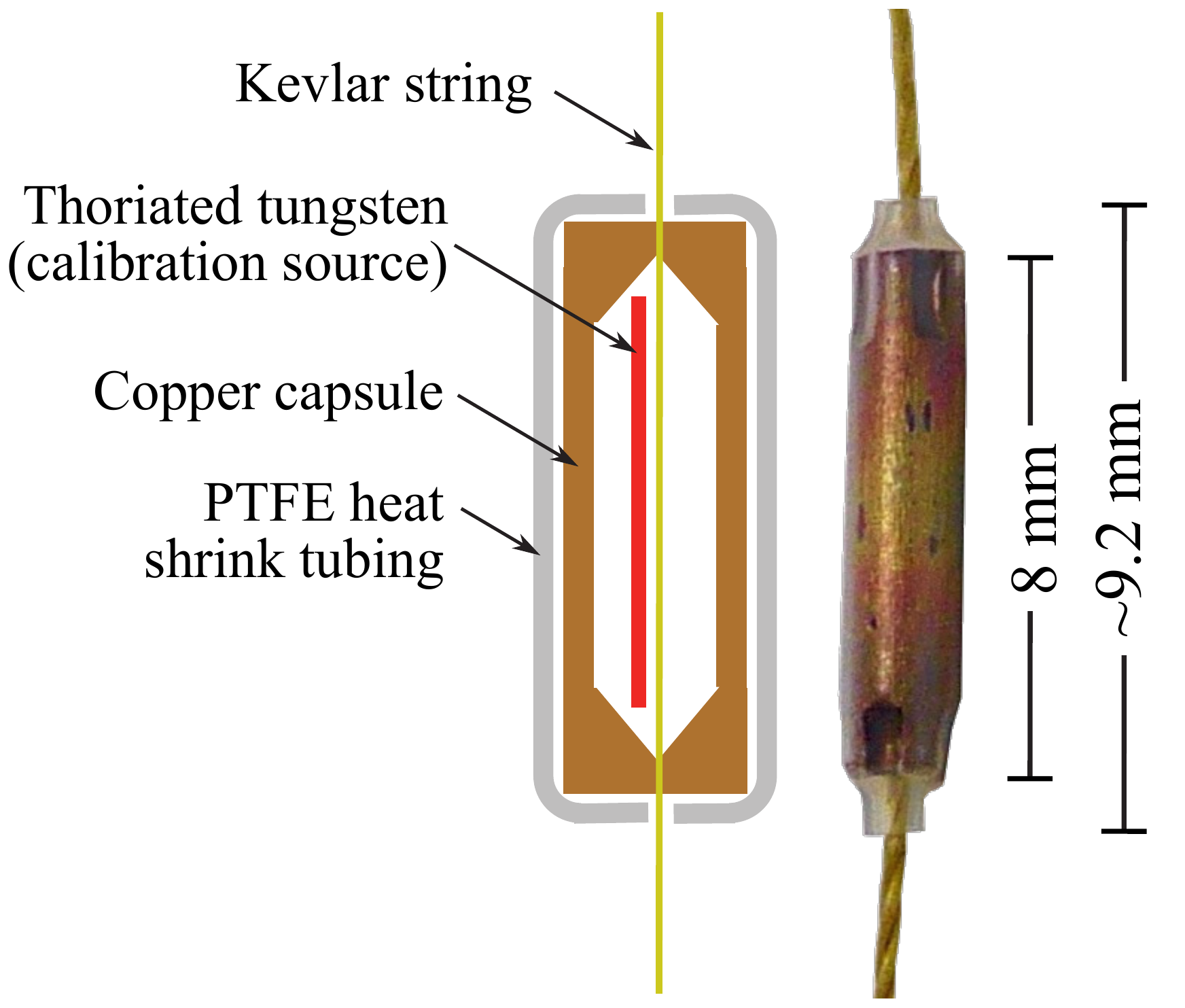}\label{fig:source_carrierA}}
\quad
\subfloat[][]{\includegraphics[height=2.7in]{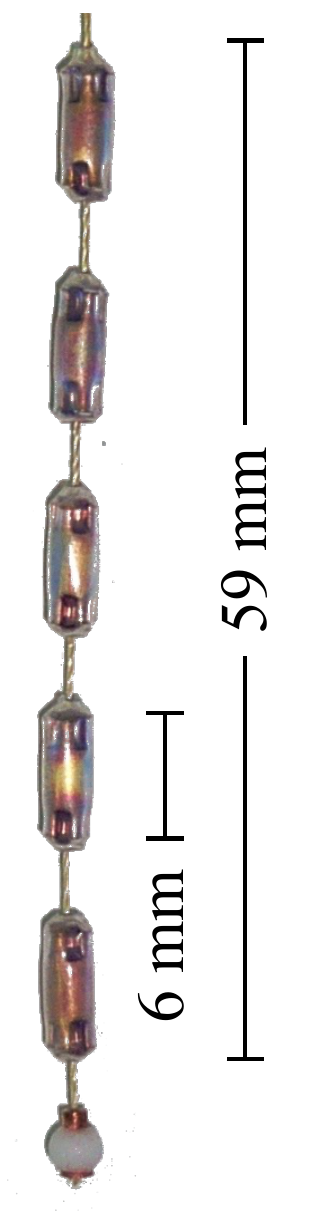}\label{fig:source_carrierB}}
\end{center}
\caption{(a) Schematic and photograph of an assembled source capsule. (b) Photograph of five heavier bottom capsules and the PTFE guide ball at the bottom of a source string.}
\label{fig:source_carrier}
\end{figure}

After the cut source wires are prepared, we place them inside small copper capsules (8.0~mm in length, 1.6~mm in diameter) that are crimped onto Kevlar\footnote{Kevlar is a registered trademark of E.I. du Pont de Nemours and Company.} strings. The capsule pitch along the string is 2.9~mm. This design allows the source strings to be flexible as they pass through the bends of the guide tubes in the cryostat, and the use of continuous strings minimizes the risk of capsules detaching from the source strings inside the cryostat. PTFE heat-shrink tubing around the capsules reduces friction and covers any edges on the source capsules that could impede their smooth motion inside the guide tubes. For the strings, we use Kevlar coated in PTFE\footnote{W.F. Lake Corporation. PTFE Coated Aramid Thread (R722-70). \mbox{\url{http://www.wflake.com}}} because of its high tensile strength and low coefficient of friction. Kevlar's low thermal conductivity at cryogenic temperatures~\cite{Ventura:2000kl} limits heat flow along the length of the string, and the PTFE coating reduces the fraying of the Kevlar string over time. The string has a diameter of 0.25~mm and a rated tensile strength of 110~N over the diameter of the string~\cite{wflakespects}. The capsule crimps are at least as strong; we are unable to slide the capsules along the string with a force less than 110~N, and any greater force breaks the string. A schematic diagram and a photograph of a source capsule are shown in \mbox{\autoref{fig:source_carrierA}}.

The source strings are lowered under their own weight as they are deployed from the top of the cryostat. To this end, we add eight larger and heavier copper capsules (6.4~mm in length, 3.2~mm in diameter, 0.4~g each) to the bottom of each string; these capsules are also loaded with source wire. In addition, we attach a small PTFE ball to the bottom of each string to help the strings enter the guide tubes inside the cryostat (see \mbox{\autoref{fig:source_carrierB}}).

\begin{figure}
\begin{center}
\subfloat[][]{\includegraphics[height=2.4in]{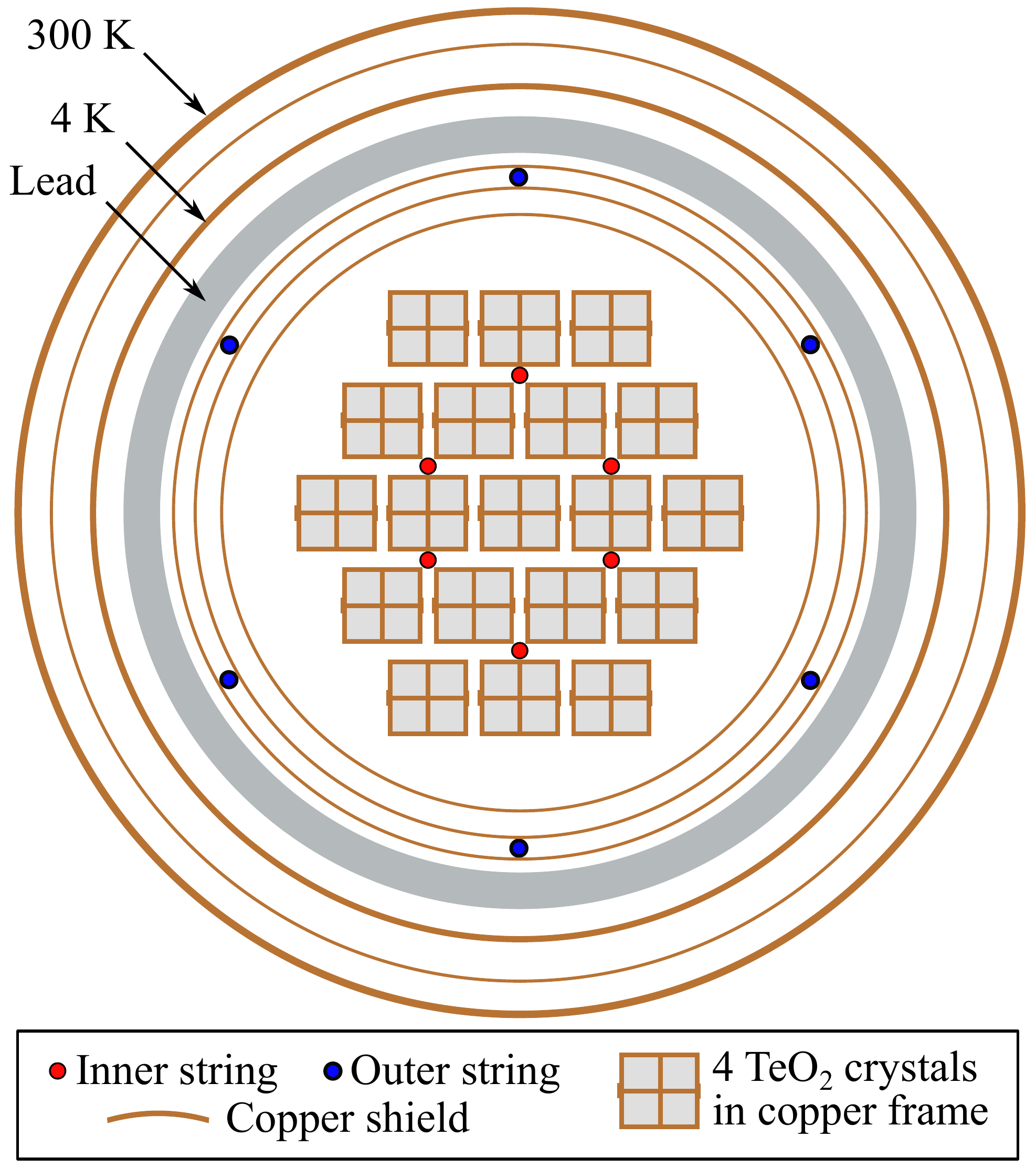}\label{fig:strings_and_towersA}}
\quad
\subfloat[][]{\includegraphics[height=2.2in]{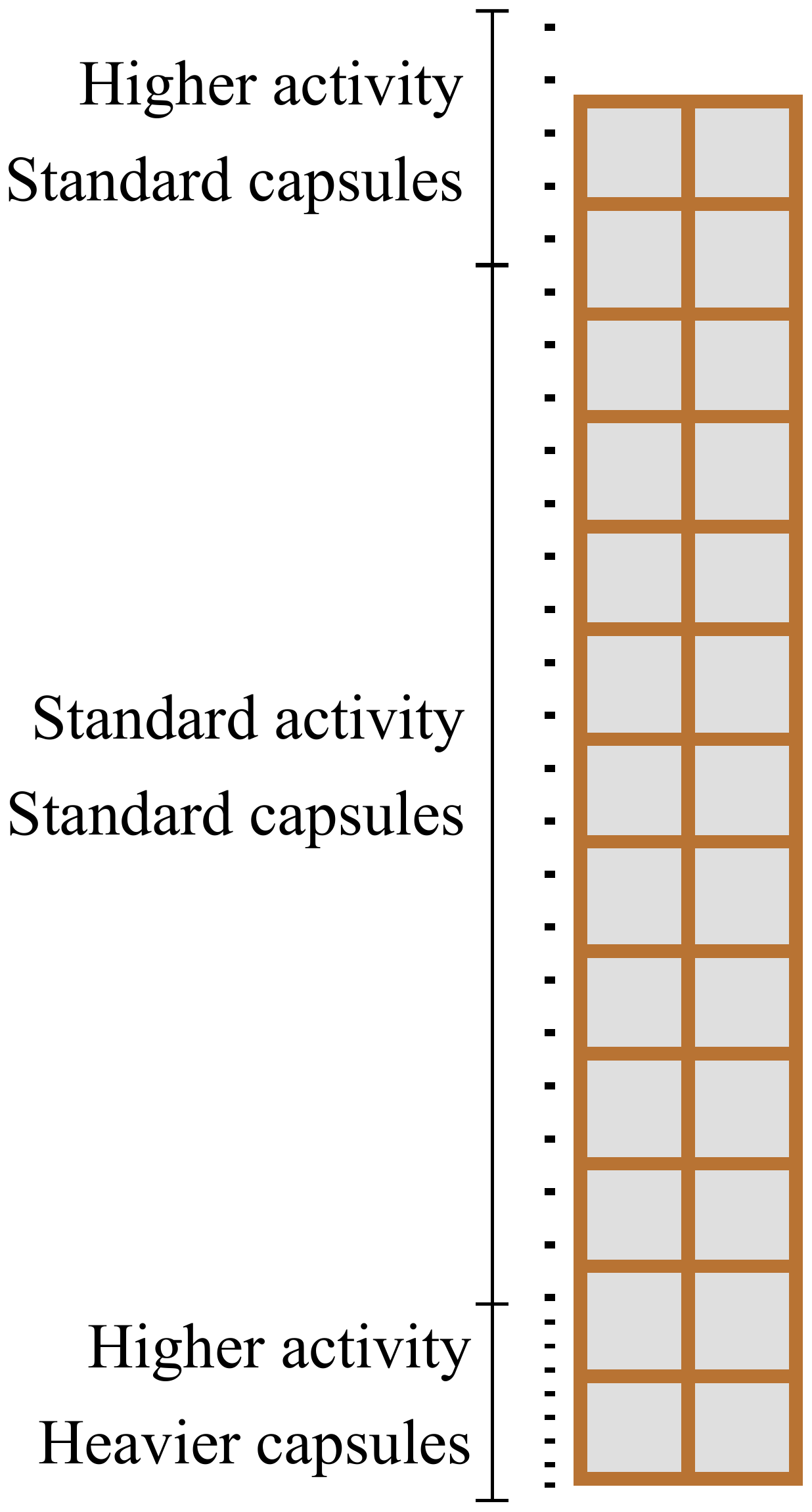}\label{fig:strings_and_towersB}}
\end{center}
\caption{(a) Top-down cross-sectional view of the 19 detector towers, showing the locations of the source strings in their calibration positions. The outer strings (19.4~Bq of $^{228}$Th each) are shown in blue; the inner strings (3.6~Bq each) are shown in red. The circles represent the cryostat vessels, which are identified as follows, starting from the outside: 300~K, 40~K, 4~K, lateral lead shielding, 600~mK, 50~mK, and 10~mK. (b) The height of the source capsules with respect to the towers, in their deployment position. Capsule size not to scale. The outer strings do not have the upper-most capsule in this figure.}
\label{fig:strings_and_towers}
\end{figure}

There are 12 deployment positions for the source strings. The distribution of the source capsules and intensities along each string matches the height of the detector towers and is optimized for event rate uniformity. The six inner strings, with sources distributed over 83~cm, are guided into the detector area and placed among the detector towers to irradiate the innermost towers. The six outer strings, with sources distributed over 80~cm, are allowed to hang outside the detector region, though still inside the cryostat's lateral lead shielding, to provide additional gamma rays to the outer towers. The locations of the calibration strings in their deployed positions with respect to the detector towers are shown in \mbox{\autoref{fig:strings_and_towers}}.

\begin{figure}
\begin{center}\includegraphics[width=3in]{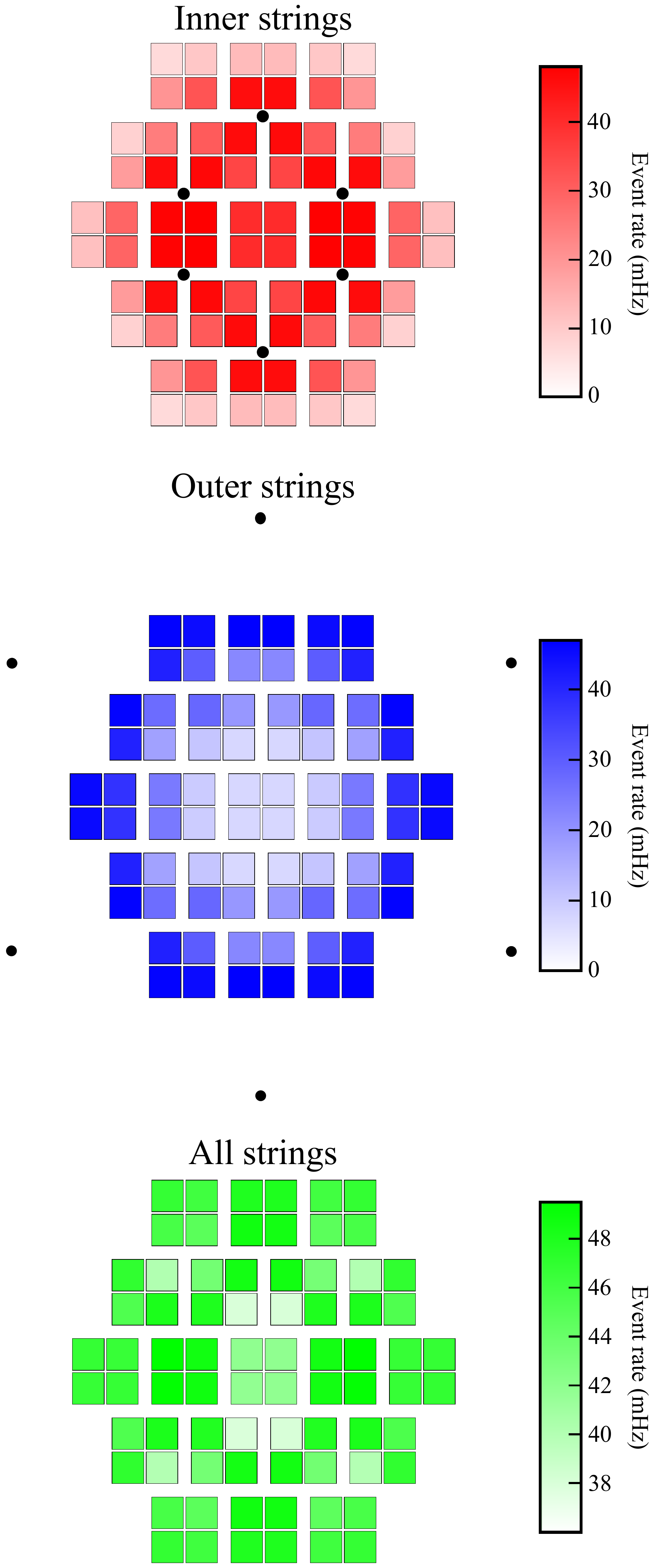}\end{center}
\caption{The simulated event rates per bolometer due to the inner strings alone (red), the outer strings alone (blue), and all strings (green), averaged over each column of 13 crystals. The total effective event rate is less than the sum of those from the inner strings and the outer strings as a result of pile-up rejection. (Color online.)}
\label{fig:rate_per_column}
\end{figure}

In high-rate calibration tests with CUORE-like crystals, we determined that the optimal trigger rate for calibration is below 150~mHz per detector~\cite{Ejzak:2013tm}. Above this rate, pile-up effects become dominant and the energy resolution is degraded. Thus, guided by Geant4-based Monte Carlo simulations~\cite{Agostinelli:2003fg} of the calibration sources in the CUORE cryostat, we selected summed source activities of 3.6~Bq of $^{228}$Th for each of the six inner strings and 19.4~Bq for each of the outer strings; this configuration results in an average trigger rate of approximately 100~mHz per bolometer. The simulated average event rates of the bolometers in each column of 13 crystals due to the inner strings alone, the outer strings alone, and all strings together are shown in \mbox{\autoref{fig:rate_per_column}}. The effective event rate per bolometer from all source strings, after pile-up rejection, is significantly less than the 100~mHz trigger rate; for this simulation, we required that the times since the previous trigger and until the next trigger on the same bolometer must be greater than 3.1 s and 4.0 s, respectively, as in \mbox{CUORE-0}~\cite{Alfonso:2015vk}.

\begin{figure}
\begin{center}\includegraphics[width=3.45in]{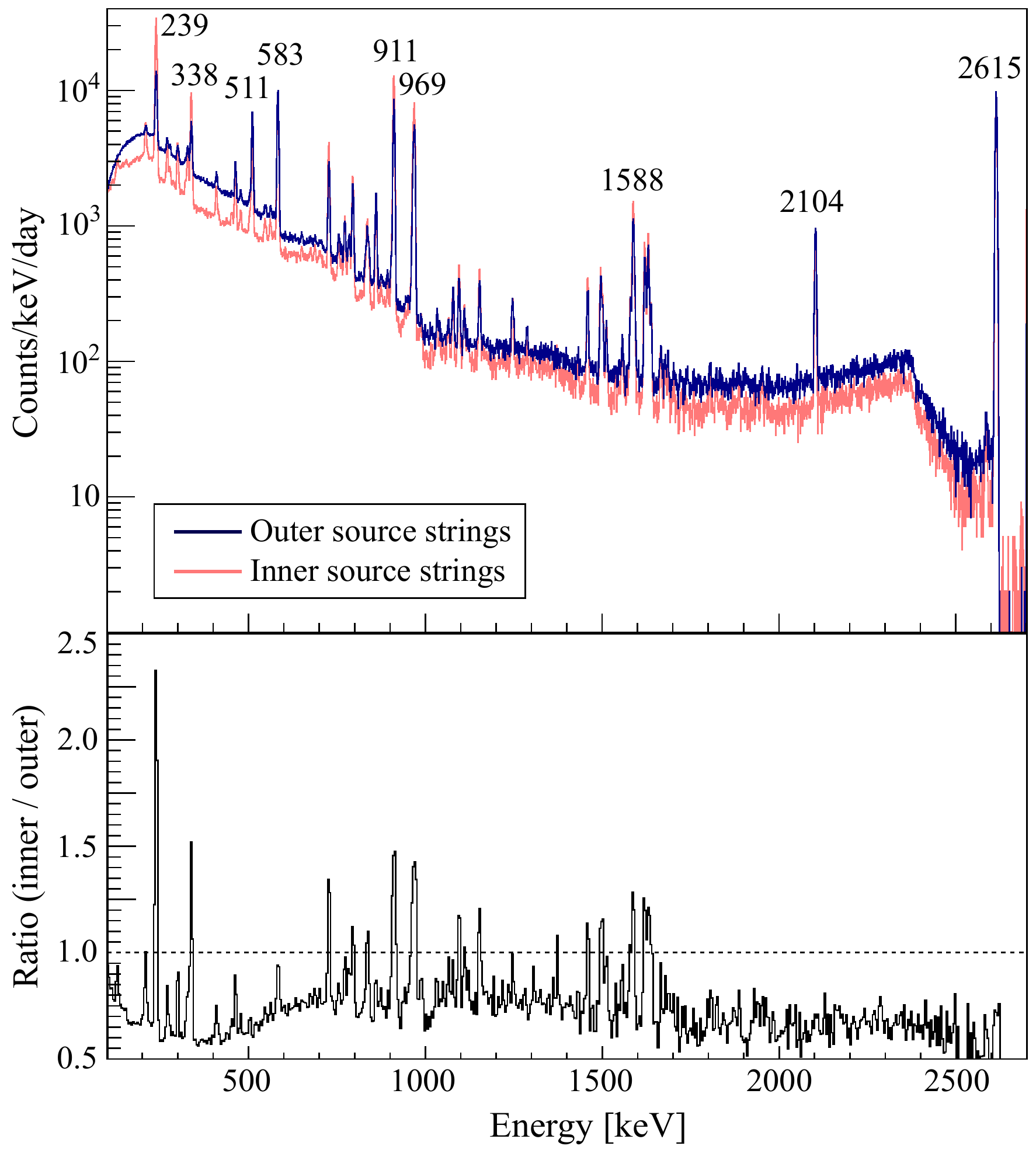}\end{center}
\caption{Top: A simulated CUORE calibration spectrum, summed over all channels. The spectra produced by the inner strings and outer strings are separated and overlaid, and the energies of important lines for calibration are labeled in units of keV. Bottom: The ratio of the counts due to the inner strings divided by the counts due to the outer strings. The outer strings have a lower peak-height-to-background ratio, particularly at lower energies, because of the presence of the copper vessels between the sources and the detectors. The ratio above 2615~keV is not shown because of low statistics.}
\label{fig:simulated_spectrum}
\end{figure}

\sloppy
The simulated calibration spectra produced by the sources are shown in \mbox{\autoref{fig:simulated_spectrum}}. The source activity is divided among the 33 or 34 capsules on each outer or inner string, respectively, with the activity at the top and bottom of the capsule region of the strings higher than that in the middle to compensate for solid-angle effects. The capsules with increased activity are distributed over $\sim$25\% of the string length but account for $\sim$50\% of the total activity. This non-uniform activity distribution was adopted because the cryostat geometry prevents the simple extension of the active source length beyond the top and bottom of the towers. Calibration sources cannot be deployed below the bottom of the bolometer towers because of the presence of the \mbox{10-mK} cryostat vessel surrounding the towers. Above the towers, the lead shielding would block the gamma rays from these low-activity calibration sources (as shown in \mbox{\autoref{fig:cryostat}}).

\fussy

\subsection{Motion control and monitoring hardware}
\label{sec:motion_control}
Before and after each calibration period, the source strings are wound on spools above the cryostat, at room temperature. These spools are attached to stepper motors that turn to deploy the strings into or extract them from the cryostat. Four stainless steel enclosures (``motion boxes'') contain the 12 source string spools and the motors that drive them; each motion box is equipped with three motors and thus controls three strings. The motion boxes can be pumped down to vacuum or vented without affecting the cryostat vacuum, allowing the replacement of any source strings as necessary during the operation of CUORE. Each motor and spool in the motion box is instrumented with several sensors to ensure the fail-safe operation of the system inside the cryostat. Many of these sensors provide redundant information to mitigate the risk of the strings being in an unknown position in the cryostat or breaking as a result of excessive tension or other exceptional circumstances.

\begin{figure}
\begin{center}\includegraphics[width=3.45in]{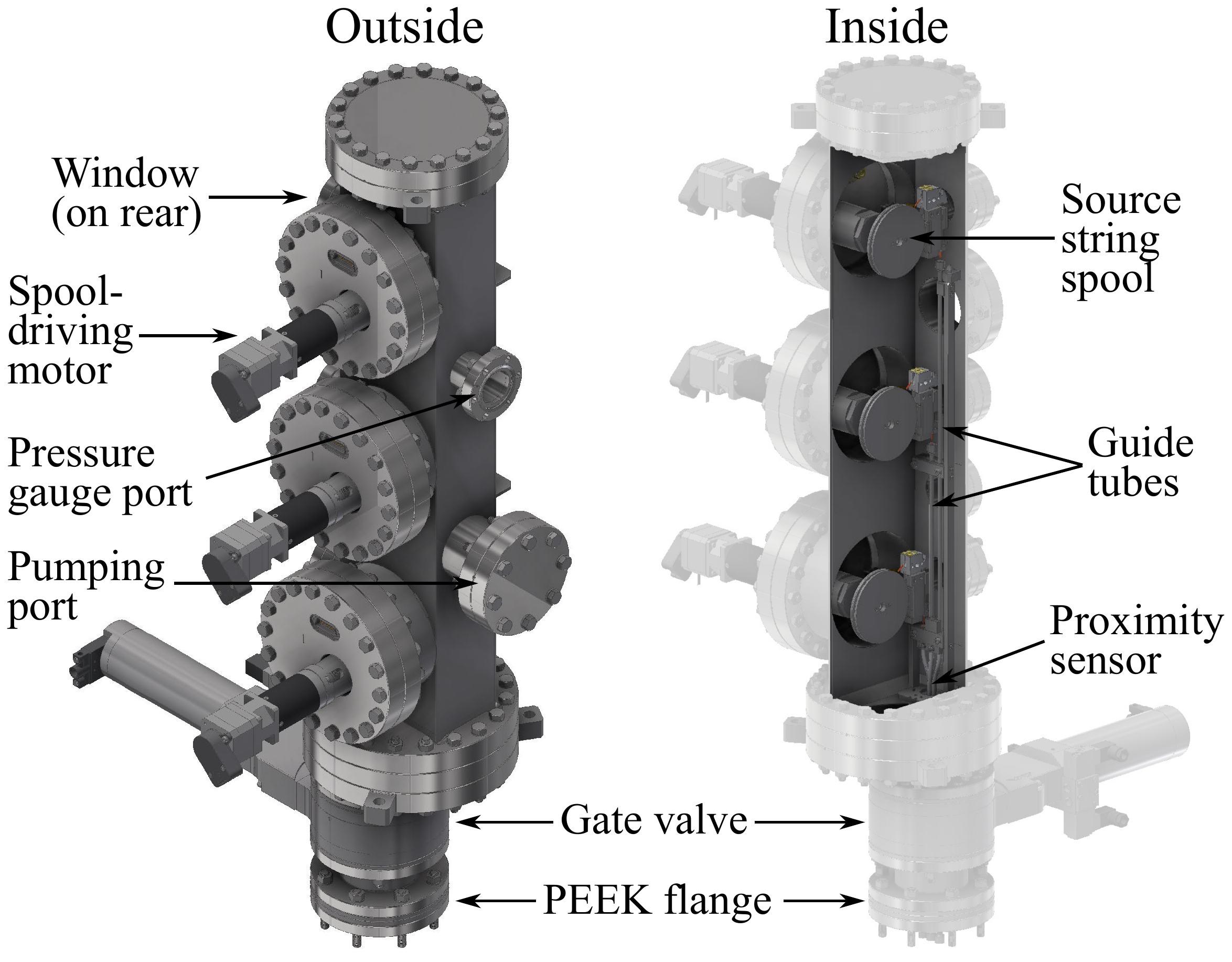}\end{center}
\caption{Rendering of a single motion box, which controls three strings. Four motion boxes are mounted above the 300-K plate of the cryostat.}
\label{fig:motion_box}
\end{figure}

The motion boxes are stainless steel weldments that are mounted to gate valves on top of the \mbox{300-K} plate of the cryostat. The main volume of a motion box is a welded box with dimensions of $126\times 79\times607~\mbox{mm}^3$. Along one face are three 6.75'' CF flanges (125~mm in inner diameter) that protrude from this box. Mated with these are three CF flanges that hold the motors, sensors, and spools containing the source strings. The flanges are aligned such that the source strings on each spool feed directly into vertical guide tubes below. Opposite the motors in the motion box are three glass viewports, which we can uncover to observe the movement of a string while it is under vacuum, if necessary. The sides of the motion box contain two additional CF flanges: one hosts a vacuum pressure gauge, and the other is for the vacuum pumping line. A rendering of a motion box is shown in \mbox{\autoref{fig:motion_box}}.

The motion boxes are sealed to the gate valves below them with fluoroelastomer O-rings. Below the gate valves, polyetheretherketone (PEEK) spacers electrically isolate the motion boxes, and thus the stepper motors, from the cryostat. The gate valves and PEEK spacers are also sealed to the cryostat and to each other with fluoroelastomer O-rings, under a compression force provided by threaded stainless steel rods and nuts that are electrically insulated from the cryostat plate by nylon sleeves and washers.

\begin{figure}
\begin{center}\includegraphics[width=3.45in]{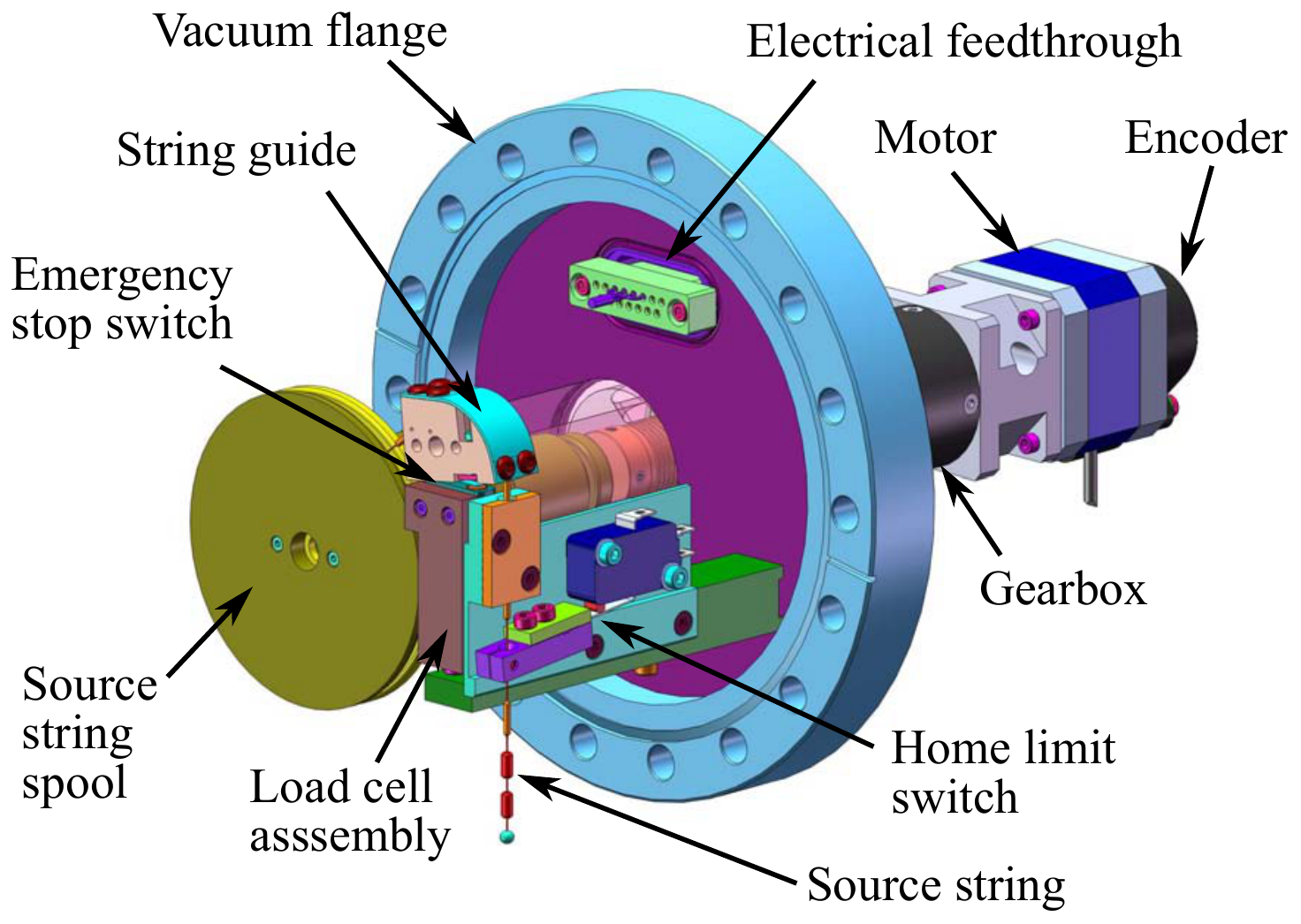}\end{center}
\caption{False-color diagram of the motor and spool controlling a single source string.}
\label{fig:drive_spool}
\end{figure}

Rotational motion is transmitted from the motor shaft outside the motion box vacuum to the source string spool through a rotary motion feedthrough\footnote{MDC Vacuum Products. Direct Drive Rotary Motion Feedthrough (652000). \mbox{\url{http://www.mdcvacuum.com}}}. This direct drive feedthrough includes a fluoroelastomer seal around the shaft; as the shaft rotates, the pressure of the fluoroelastomer against the shaft maintains the vacuum inside the motion box. The feedthrough is attached to the motor shaft outside the vacuum with a bellows coupling. A diagram of the motor, spool, and feedthrough mounted on a CF flange is shown in \mbox{\autoref{fig:drive_spool}}.

The stepper motors that raise and lower the strings\footnote{Kollmorgen. CT Series Stepper Motor (CTP10ELF10MMA00). \mbox{\url{http://www.kollmorgen.com}}} have a 1.8$^\circ$ step angle and are connected to a 320:1 planetary gearbox (gear reducer)\footnote{Anaheim Automation. GBPN-064x-FS Series Planetary Gearbox (GBPN-0403-320-AA171-197). \mbox{\url{http://www.anaheimautomation.com}}}. Thus, $64\,000$ steps of the stepper motor result in one full rotation of the source string spool. The strings can be moved smoothly at speeds ranging from below 5~mm/minute to above 500~mm/minute.

Information regarding the position and velocity of the strings in the cryostat is provided by a rotary encoder on each motor\footnote{US Digital. E5 Optical Kit Encoder (E5-1000-197-IE-D-D-G-B). \mbox{\url{http://www.usdigital.com}}}. This 1000 cycles-per-revolution optical incremental encoder, combined with the gear reducer installed on each motor, yields a position resolution of approximately 1~$\mu$m, which is particularly useful for velocity measurements during the slowest parts of the deployment and extraction. By measuring the motion of the motor shaft, the encoders verify that the string is actually moving inside the cryostat and can alert the system operator of non-operational or malfunctioning motors or motor controllers.

As a string is lowered into the cryostat, we continuously monitor the string tension using a load cell\footnote{Strain Measurement Devices. Miniature Platform Load Cell S251 (SMD2387-05-020). \mbox{\url{http://www.smdsensors.com}}}. The load cell has a 3000-$\Omega$ thin-film strain gauge bridge and a range of 0 to 2000~g. A 1-$\mu$F capacitor is placed across the signal leads from the load cell to reduce high-frequency noise from the nearby stepper motors. The load cell is connected to an in-line amplifier\footnote{Honeywell. Bridge Based Sensor In-Line Amplifier (060-6827-04). \mbox{\url{https://measurementsensors.honeywell.com}}} that is read out by a 12-bit analog-to-digital converter (ADC).  The result is a tension reading that permits the detection of changes at a level of 5~mN or less. The commissioning of a source string path involves repeatedly raising and lowering a string through that path to determine the standard profile of the string tension as a function of the string position and direction. During all subsequent deployments, we ensure that the data from each load cell are consistent with its load cell profile; if there are any sustained deviations, the string is stopped, withdrawn, and redeployed. In testing with intentionally misaligned guide tubes, we have found that this method is able to reliably detect when a string has failed to enter a guide tube during deployment.

Each string enters and leaves its spool horizontally and passes over a PTFE string guide, which is mounted on the lever actuator of a micro switch. This guides the string into the opening of the guide tube below during the string deployment and, when extracting the string, triggers the motion system to stop if there is significant tension on the string. In normal operation, this switch should never be triggered. After the string passes over the PTFE guide, it passes through an aperture in the lever actuator of a second micro switch that we use as a home position indicator. As the string is withdrawn from the cryostat, when the first larger capsule at the bottom of the string hits this lever actuator, it triggers the micro switch.

Near the bottom of each motion box, an inductive proximity sensor\footnote{Proxitron. Inductive Ring Sensor (IKVS-010.23-G-S4). \mbox{\url{http://proxitron.de}}} detects and counts the copper source capsules entering and leaving the cryostat. We use this signal to ensure that all of the capsules have unspooled successfully and have entered the cryostat, and after calibration has been completed, we use it as an additional verification that all of the source capsules are fully withdrawn from the cryostat.

Our string position uncertainty is dominated by the effects of spooling and unspooling the source strings, and the source capsules in particular. To remove this uncertainty, we redefine the zero position of each string as its final source capsule enters the proximity sensor during the string deployment. This occurs before the first source capsules cross the \mbox{4-K} plate of the cryostat and results in a final position uncertainty of approximately 1~mm in the lower regions of the cryostat.

All of the motors, motion box sensors, and temperature and pressure monitors are read and controlled by a dedicated server, which enables us to perform the calibration sequence and monitor its status remotely and automatically (see \mbox{\autoref{sec:control_electronics}}).

\subsection{Source string guide tubes}
\label{sec:guide_tubes}
\begin{figure}
\begin{center}\includegraphics[height=4.5in]{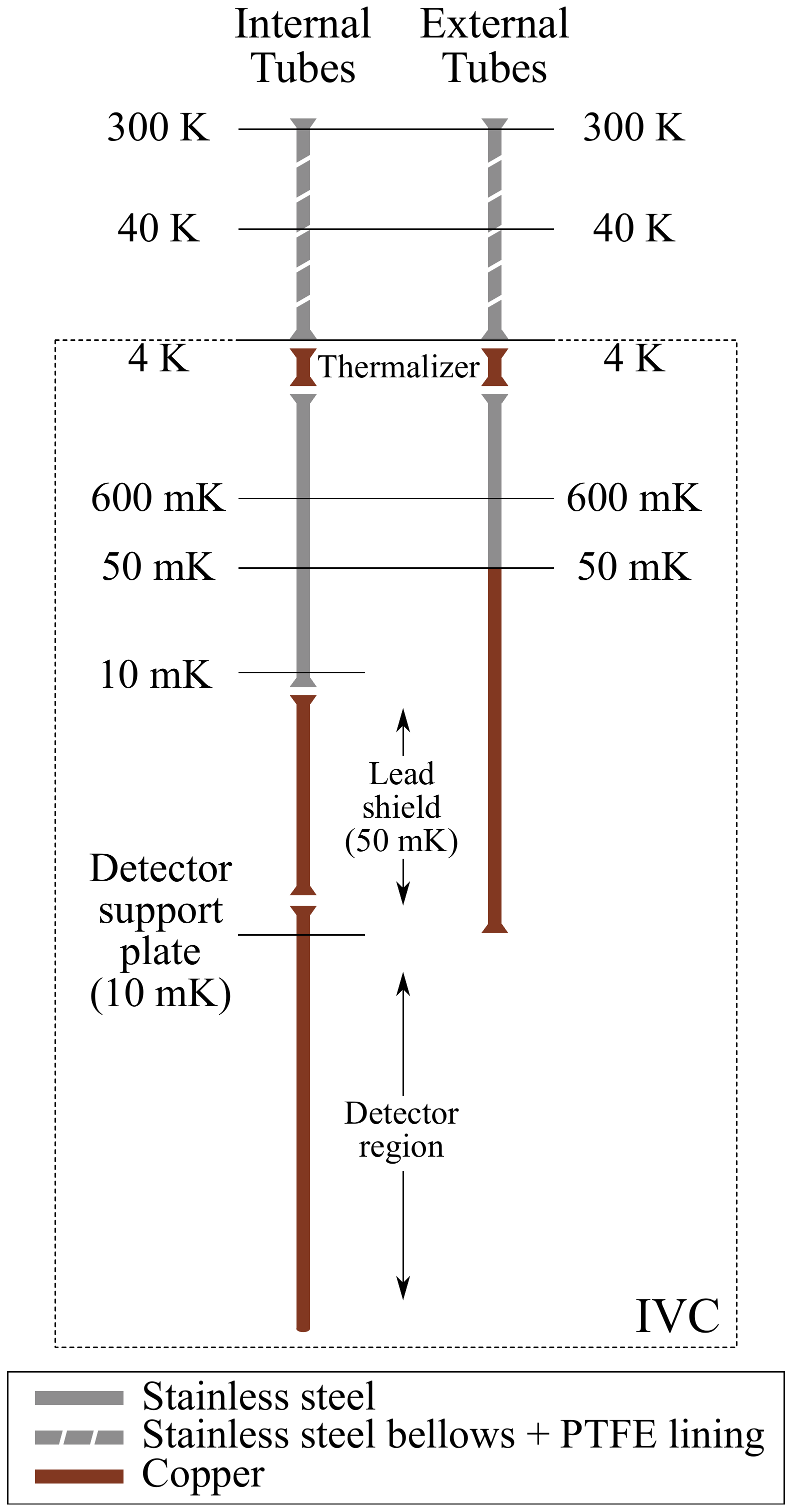}\end{center}
\caption{Schematic of the DCS guide tubes in the CUORE cryostat. All tubes are fully thermalized to the cryostat plates that they cross. The region surrounded by the dotted line represents the inner vacuum chamber. The thermalizer at 4~K is discussed in \mbox{\autoref{sec:thermalizer}}.}
\label{fig:thermal_coupling}
\end{figure}

The source strings pass through several different guide tubes en route to the detector region at 10~mK. Each source string has its own set of guide tubes; the strings are thus isolated from each other at all points to avoid any possibility of the strings becoming entangled or stuck. A schematic of the layout of the DCS guide tubes in the CUORE cryostat is shown in \mbox{\autoref{fig:thermal_coupling}}.

\sloppy
The DCS is fully integrated with a complex cryostat, and as such, the system's heat load must not exceed the available cooling power at any thermal stage. The guide tubes in the warmer regions are made of stainless steel, and the source strings are constructed from Kevlar; the low thermal conductivity of these materials limits the heat transfer between stages. During the deployment of a string into the cryostat, the source capsules are cooled before entering the most sensitive parts of the cryostat to avoid placing a large heat load on the colder cryostat stages. During string extraction, the friction between the strings and the guide tubes accounts for the majority of the heat load; in addition to using PTFE-coated capsules and source strings and highly polished guide tubes, we minimize the frictional heat load by moving the strings very slowly ($\sim$10~mm/minute) in the coldest regions of the cryostat.

\fussy
The CUORE cryostat has two vacuum chambers: the inner vacuum chamber (IVC) and the outer vacuum chamber (OVC). The walls of the IVC are the \mbox{4-K} cryostat plate and vessel, whereas the room-temperature (300~K) plate and vessel form the boundaries of the OVC. The motion boxes and all string paths open directly into the IVC. In practice, this means that the motion boxes and guide tubes above the \mbox{4-K} cryostat plate must be vacuum-tight, but those below the \mbox{4-K} plate can be open to the IVC around them.

\begin{figure}
\begin{center}\includegraphics[width=2.5in]{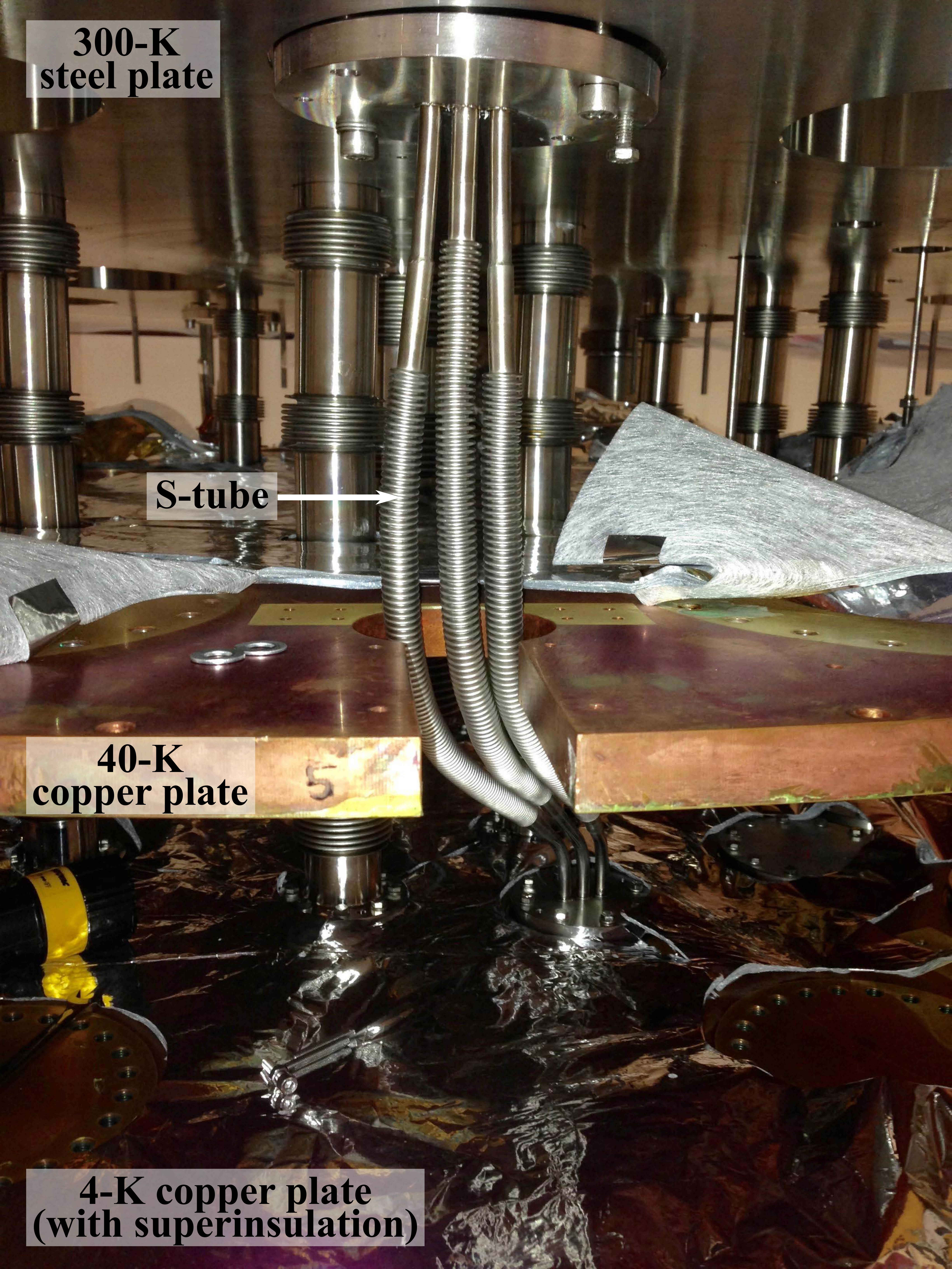}\end{center}
\caption{An S-tube assembly being installed in the cryostat. Each such assembly contains three string paths and connects the \mbox{300-K} steel plate (top) to the \mbox{4-K} copper plate (bottom). The tubes pass through a hole in the \mbox{40-K} plate, where they are thermalized to the plate with a clamp (not shown).}
\label{fig:s_tube}
\end{figure}

As the source strings are lowered below the gate valves in the motion boxes, the first guide tubes they encounter are S-shaped tubes (``S~tubes'') that bring the strings from room temperature to the \mbox{4-K} cryostat plate (see \mbox{\autoref{fig:s_tube}}). These S~tubes have a coaxial design, with stainless steel vacuum-tight thin-wall formed bellows surrounding fluorinated ethylene propylene (FEP) tubing with an inner diameter of 4~mm. This design accommodates the relative motion of the \mbox{300-K}, \mbox{40-K}, and \mbox{4-K} plates that arises from the cryostat suspension system and from the thermal contraction of the cryostat as it cools. The stainless steel bellows provide structural integrity with low thermal conductivity, and the FEP tubing helps to minimize the friction that the strings experience as they move through the S~tubes. The bellows maintain a temperature gradient between 300~K and 4~K that helps to cool down the string as it is lowered. The FEP tubing is also flared at the ends using a thermal forming process; these flares act as funnels, helping the strings to enter the S~tubes without becoming caught on any edges. The S~tubes pass through slots in the \mbox{40-K} cryostat plate, where they are thermalized to the plate with copper braids squeezed around the bellows with an indium shim. The assembly is vacuum sealed to the \mbox{300-K} cryostat plate with a fluoroelastomer O-ring and to the \mbox{4-K} plate with indium.

At the bottom of the bellows, the strings enter a thermalizer that mechanically squeezes the source capsules on the string to cool them down to 4~K. We discuss this thermalization mechanism further in \mbox{\autoref{sec:thermalizer}}.

\begin{figure}
\begin{center}\includegraphics[width=2.5in]{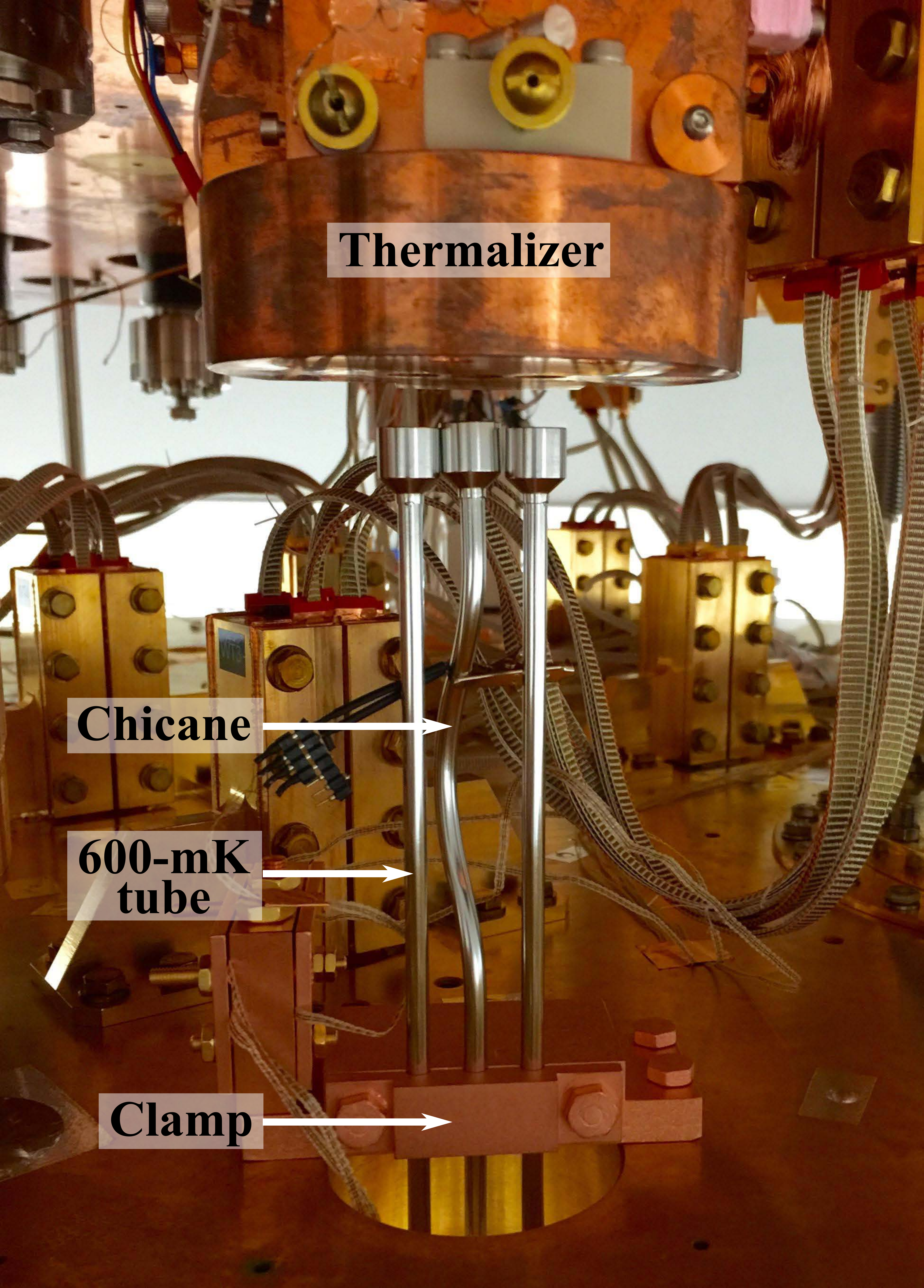}\end{center}
\caption{Photograph of the \mbox{600-mK} guide tubes installed below a thermalizer. There is a visible chicane in the center tube. The tubes are thermalized to the \mbox{600-mK} plate via the copper clamp at the bottom of the image. The copper clamp on the left thermalizes the thermometer wires attached to the tubes.}
\label{fig:600mk_tube_photo}
\end{figure}

Below the \mbox{4-K} thermalizer, the strings enter stainless steel guide tubes. These tubes have an inner diameter of 4~mm and a wall thickness of 0.38~mm. They are mechanically and thermally anchored to the \mbox{600-mK} cryostat plate with copper clamps and separated from the thermalizers by a gap of approximately 1~cm. This gap limits thermal conduction between the \mbox{4-K} plate, to which the thermalizer is anchored, and the \mbox{600-mK} plate; however, the tubes do maintain a small temperature gradient above 600~mK because of radiation from the nearby thermalizer. On both sides of the gap, there are funnels to guide each string into the next tube. Above the \mbox{600-mK} plate, the six tubes that are on inner guide tube paths each have additional bends in the shape of a chicane to improve the contact between the source capsules and the walls of the tube and thus to improve the thermalization of the capsules (see \mbox{\autoref{fig:600mk_tube_photo}}).

\sloppy
The guide tubes that contain inner source strings are made of stainless steel until they reach the \mbox{10-mK} plate. They are split at the \mbox{600-mK} plate for installation into the cryostat, but the upper and lower portions are clamped together with no gap between them. In addition to being thermalized to the \mbox{600-mK} plate by copper clamps, they are also similarly thermalized to the \mbox{50-mK} and \mbox{10-mK} plates and thus maintain a temperature gradient from 600~mK to 10~mK. These guide tubes are also sloped from the \mbox{50-mK} plate to the \mbox{10-mK} plate (45--59$^\circ$ off vertical), encouraging source capsule thermalization with the walls of the tube. Below the \mbox{10-mK} plate, there is a gap, and the inner source strings then continue into copper guide tubes inside the cryostat's lead shielding. Below the lead shield, there is another gap, and the strings then enter the detector-region guide tubes. The gaps above and below the lead shield allow the detector-region guide tubes to be vibrationally isolated from the lead shielding and from the other guide tubes. The inner source strings are fully contained in copper tubes at all times when they are in the detector region. These tubes capture thermal radiation from the source strings, prevent the contamination of the detector towers by the source strings, and ensure that the sources remain in the correct position for calibration.

\fussy
All of the tubes below the \mbox{10-mK} plate are vertical to minimize the friction between the source strings and the tubes. The tubes in the detector region (6~mm in inner diameter) are also larger in diameter than those above to further reduce contact with the walls of the tubes and the resulting friction. The tubes that pass through the lead shielding and those in the detector region are composed entirely of electrolytic tough pitch copper\footnote{Aurubis. ETP-1 Oxygen-Bearing Copper (NOSV). \mbox{\url{http://www.aurubis.com/}}} subjected to tumbling, electropolishing, chemical etching and plasma etching~\cite{Alessandria:2013fs} to respect the strict background radioactivity requirements of the cryostat near the detectors.

\begin{figure}
\begin{center}\includegraphics[width=3.45in]{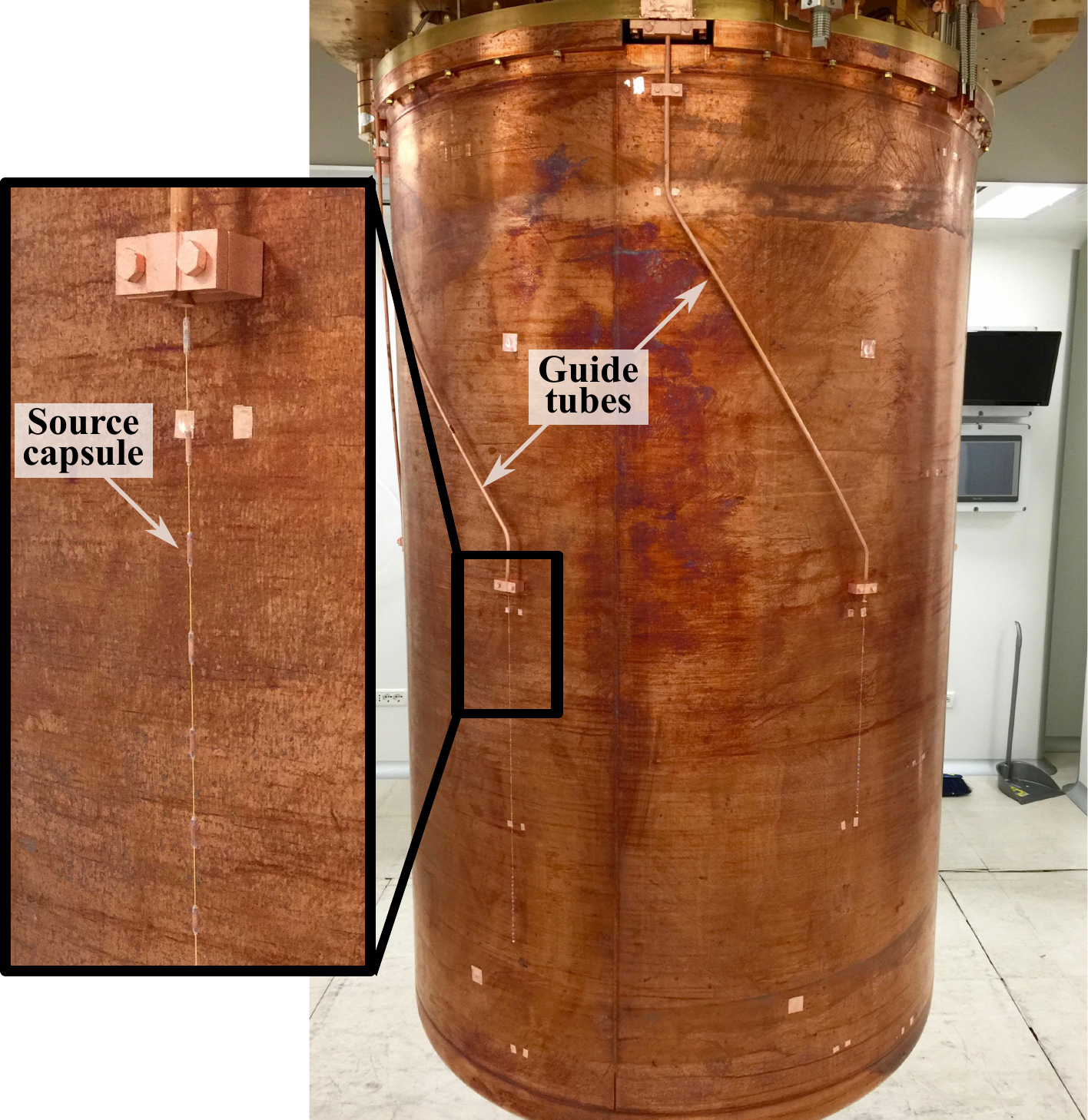}\end{center}
\caption{Helical outer guide tubes at \mbox{50-mK} installed in the CUORE cryostat. A source string is shown in a partially deployed position as it is lowered through a guide tube.}
\label{fig:helical_tubes_photo}
\end{figure}

The outer source strings continue below the \mbox{4-K} thermalizer in stainless steel tubes until they reach the \mbox{50-mK} cryostat plate. These tubes are sloped between the \mbox{600-mK} and \mbox{50-mK} plates (34$^\circ$ off vertical) and are thermally coupled to both; thus, they maintain a temperature gradient that cools the strings as they are lowered. Below the \mbox{50-mK} plate, helical oxygen-free high-thermal-conductivity (OFHC) copper guide tubes bring the strings down to the detector region (see \mbox{\autoref{fig:helical_tubes_photo}}). These tubes are thermally anchored to the outside of the \mbox{50-mK} cryostat vessel and to the \mbox{50-mK} plate, and they connect seamlessly to the stainless steel guide tubes above the plate. The slope of the helical tubes (15--46$^\circ$ off vertical) allows the strings to thermalize to 50~mK through contact with the tubes as they are lowered. Below the helical tubes, the outer strings are allowed to hang freely because they are spatially separated from the detectors by the \mbox{10-mK} and \mbox{50-mK} cryostat vessels.

\subsection{\mbox{4-K} thermalization mechanism}
\label{sec:thermalizer}
\begin{figure}
\begin{center}\includegraphics[width=3.45in]{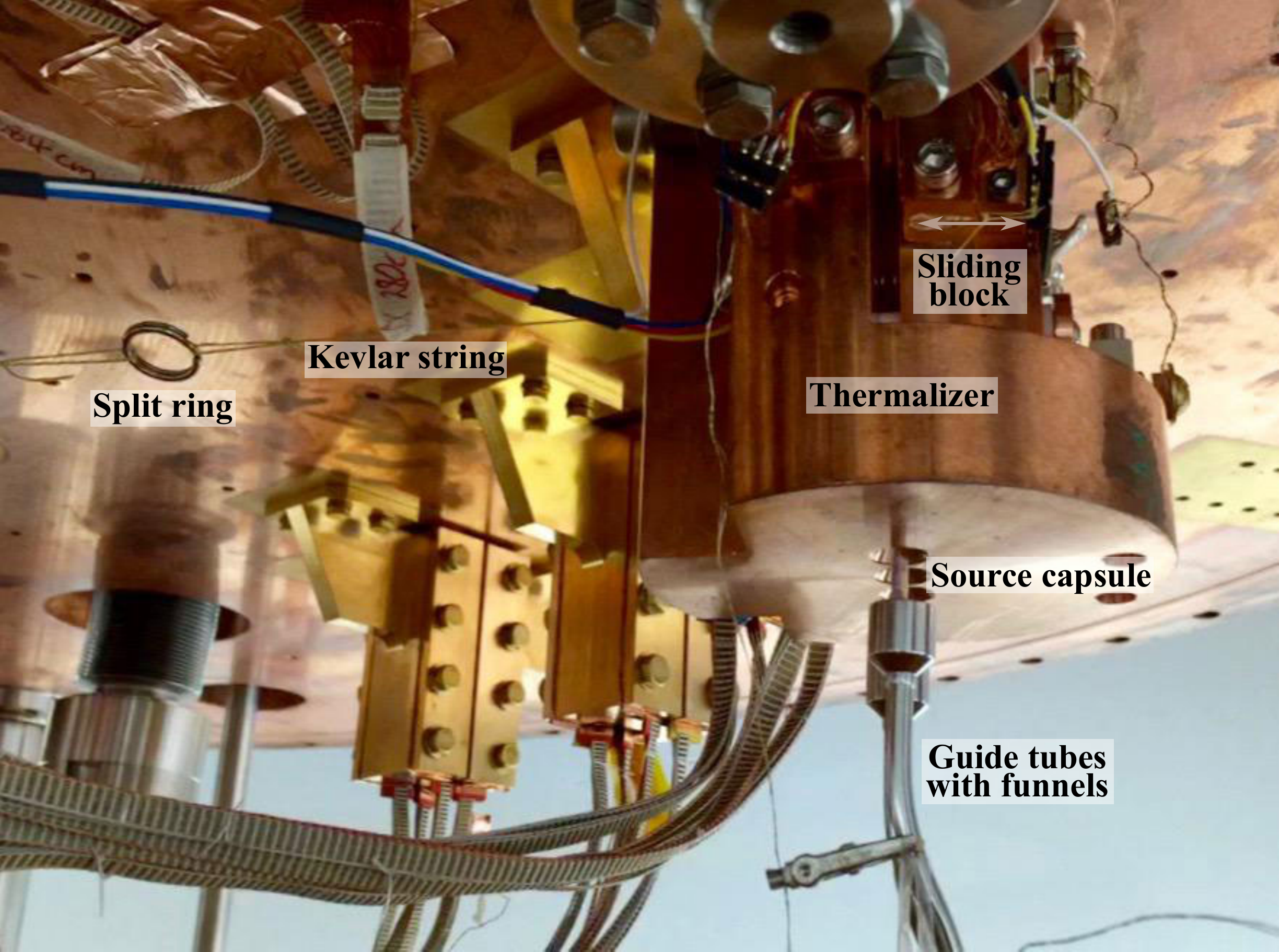}\end{center}
\caption{A thermalizer mounted in the CUORE cryostat. A capsule is visible entering the stainless steel guide tubes below the thermalizer.}
\label{fig:photo_thermalizer}
\end{figure}

One of the critical requirements of the DCS is to ensure that lowering the source strings into the cold cryostat does not change the operating temperature of the bolometers; that is, the temperature stabilization system must be able to compensate for the head load introduced by the source strings as they are lowered. It is therefore desirable to have an additional thermalization mechanism in place, as simple contact between the source strings and the guide tubes is not sufficient to cool down the source capsules efficiently. We have created such a mechanism, the \mbox{4-K} thermalizer, which makes mechanical contact with each source capsule as it is lowered into the cryostat. There are four such thermalizers in the cryostat; the three strings deployed from a single motion box pass through the same thermalizer. A photograph of a thermalizer is shown in \mbox{\autoref{fig:photo_thermalizer}}.

\begin{figure*}
\begin{center}\begin{tabular}{cc}
\includegraphics[height=3in]{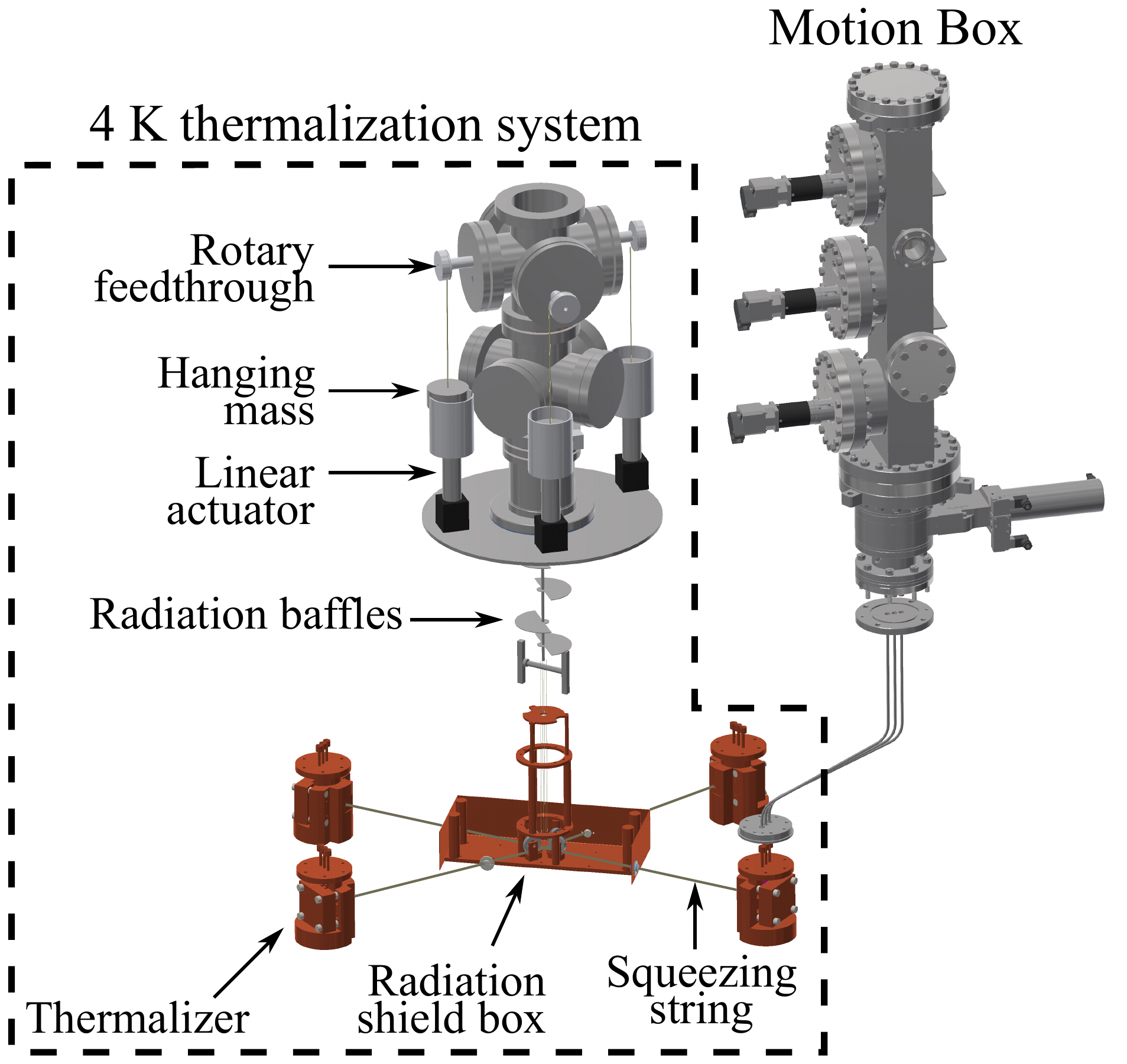} & \includegraphics[height=2.6in]{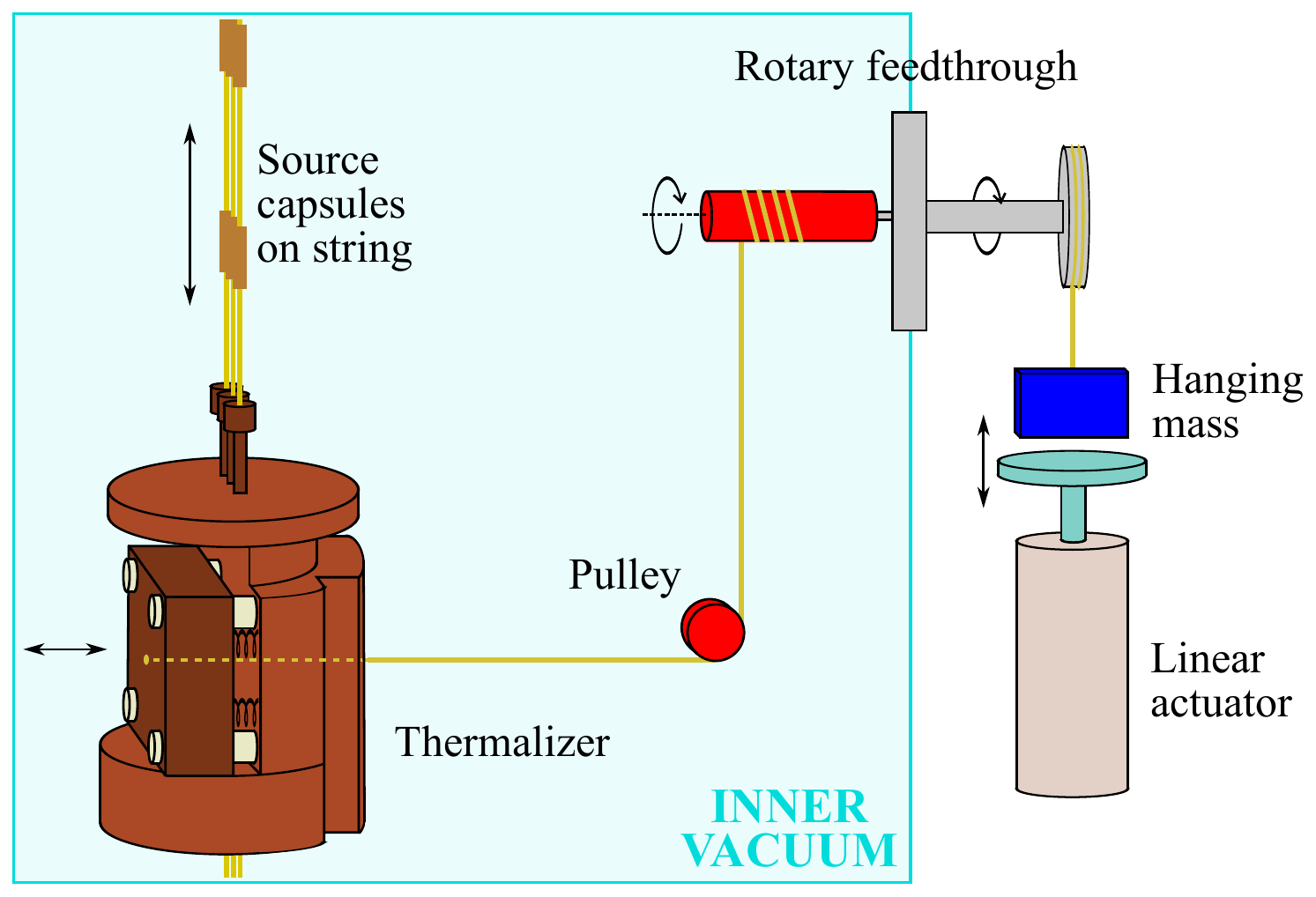}\\
(a) & (b)
\end{tabular}\end{center}
\caption{(a) Rendering of the \mbox{4-K} thermalization hardware (boxed), with a motion box and an S-tube assembly containing source strings above one of the four thermalizers. (b) A schematic of one thermalizer and its accompanying hardware.}
\label{fig:thermalizer_schematic}
\end{figure*}

Each thermalizer is well anchored, physically and thermally, to the bottom of the \mbox{4-K} plate of the cryostat. This location was chosen because it is the coldest cryostat plate at which there is a large amount of cooling power available to the DCS (see \mbox{\autoref{tab:cooling_power}}). The thermalizer is primarily composed of a fixed copper body and a sliding copper block on PTFE guides; the sliding block is held apart from the body by a spring but is thermally coupled to it with a copper braid. The sliding block is attached to a Kevlar string, and when tension is applied to this string, the block is pulled against the spring and moves closer to the fixed body of the thermalizer, squeezing any capsules that are in between. This Kevlar string is connected to another Kevlar string that runs to the shaft of a room-temperature rotary vacuum feedthrough\footnote{Ferrotec. Ferrofluidic Seal Thread Mounted Feedthrough (SS-250-SLAB). \mbox{\url{https://www.ferrotec.com}}} at the top of the cryostat via a well-polished split ring, for ease of installation. A reproducible, fixed amount of tension is applied from outside the cryostat by means of a hanging mass. To open the thermalizer, a linear actuator lifts the hanging mass to release the tension on the Kevlar string, and the springs cause the sliding block to move away from the body of the thermalizer. Each thermalizer can squeeze two normal source capsules on a single string or four of the more closely spaced heavier source capsules at the bottom of the string. A three-dimensional rendering and a schematic of the thermalization system are shown in \mbox{\autoref{fig:thermalizer_schematic}}.

All four thermalizers are controlled by individual hanging masses and linear actuators outside the vacuum and at room temperature above the cryostat. The force from each hanging mass is transmitted to the string inside the cryostat that closes the thermalizer via a rotary feedthrough, in which a hermetic ferrofluid seal around the rotating shaft maintains the vacuum inside the cryostat.

The four linear actuators are located together above the center of the cryostat, and the four Kevlar strings that cause the thermalizers to close pass through a single port on top of the cryostat (see \mbox{\autoref{fig:thermalizer_schematic}}). This port opens directly into the IVC via stainless steel tubing and bellows, and it has line-of-sight access through the \mbox{4-K} plate. The strings pass through this tubing and break out toward the four thermalizers below the \mbox{4-K} plate of the cryostat, turning 90$^\circ$ from vertical to horizontal on small pulleys. Baffles in the tubing and a small copper box around the pulleys, internally coated with polyimide film, minimize the radiation that reaches below the \mbox{4-K} stage. The strings emerge from small cutouts in the sides of the box via PTFE string guides.
 
The 4-K thermalization system is instrumented in several ways to ensure that an operating thermalizer has closed onto a capsule inside the cryostat. First, outside the cryostat, a potentiometer indicates the position of the linear actuator, with which we can verify that the hanging mass is being raised or lowered. Second, gold contact pads on either side of the sliding block of the thermalizer are grounded when the thermalizer is either fully open or fully closed, whereas they are ungrounded (floating) when the thermalizer is between these extremes. Thus, we can verify that the sliding block has moved, and we can also distinguish between squeezing on a source capsule and squeezing on nothing or only the Kevlar string; if there is a source capsule in the thermalizer, it will not close fully, and if there is no capsule in the thermalizer, it will. Finally, the sliding block of the thermalizer exhibits a measurable rise in temperature when the cold thermalizer squeezes onto a warmer source capsule, which we measure using dedicated thermometers.

In tests at 4~K, we squeezed a PTFE-encapsulated silicon diode thermometer in the cryostat with a variety of different forces to test the cooling power of the thermalizer. We observed that the cooling time decreased as the force on the capsule increased, up to approximately 32~N. Larger forces did not result in significantly improved cooling times and are more likely to deform the capsules. Thus, a force of 32~N was selected for the final installation. 

\section{Electronic control and monitoring system}
\label{sec:control_electronics}
\begin{figure*}
\begin{center}\includegraphics[width=6in]{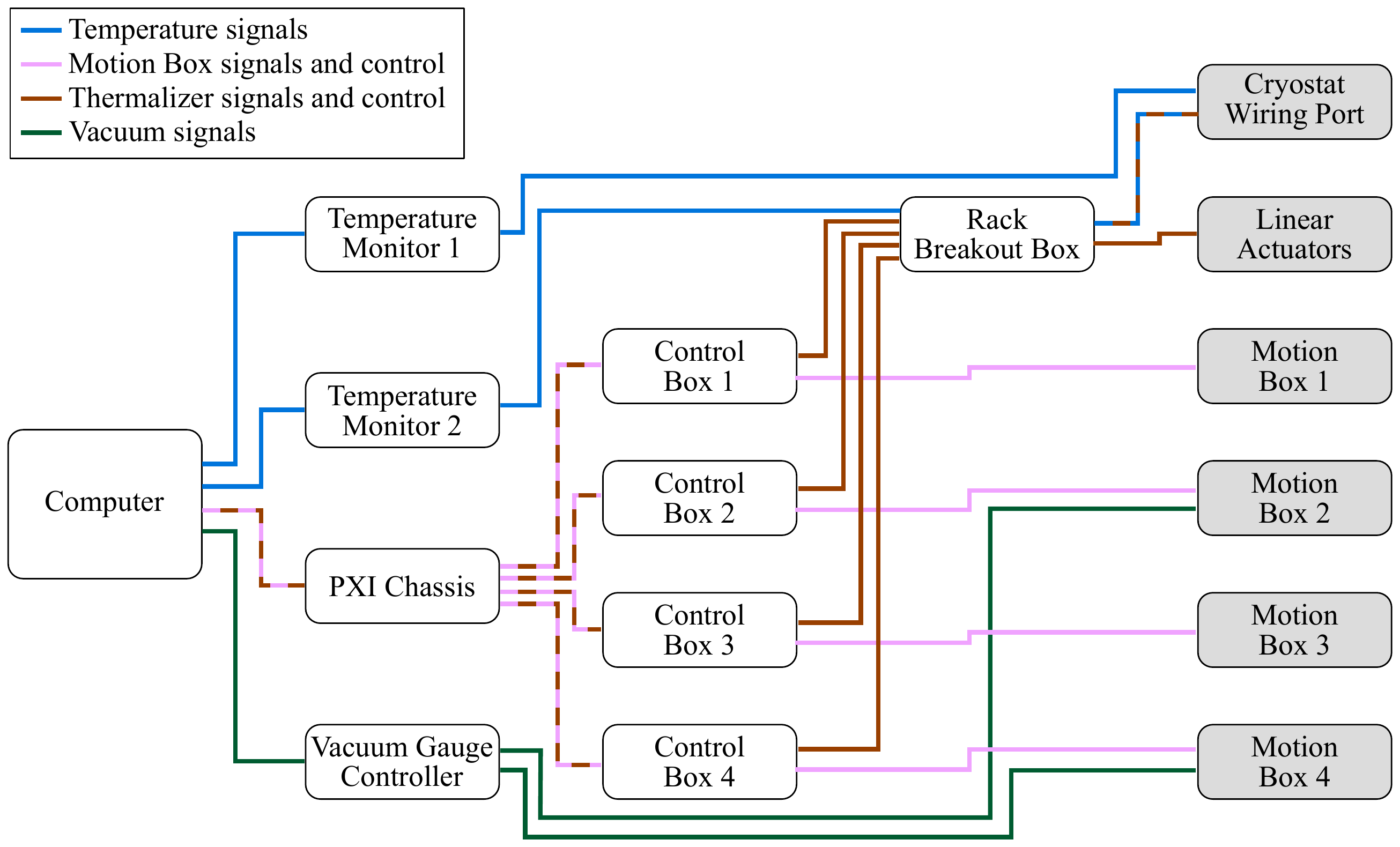}\end{center}
\caption{Overview of the DCS control system. White boxes represent components hosted in the control rack; shaded boxes represent components mounted on the \mbox{300-K} plate of the cryostat. Because all of the motion boxes are part of the same vacuum system, there is no need for an individual gauge on each.}
\label{fig:system_overview}
\end{figure*}

The DCS is a fully automated system, controlled and monitored by a dedicated server located next to the CUORE cryostat. A schematic of the DCS control system is shown in \mbox{\autoref{fig:system_overview}}.

\subsection{Hardware}
\sloppy
The server communicates directly with four pieces of hardware: two temperature monitors, a vacuum gauge controller, and a PXI chassis\footnote{National Instruments. 5-Slot PXI Chassis with Integrated MXI-Express Controller (PXI-1033). \mbox{\url{http://www.ni.com}}}. The PXI chassis contains four stepper motor controller PXI cards\footnote{National Instruments. 4-Axis Mid-Range Stepper/Servo Motion Controller (PXI-7340). \mbox{\url{http://www.ni.com}}}, each of which controls four motors and contains an integrated four-channel 12-bit ADC and 32 digital input/output lines. Each PXI card is in turn connected to a custom-designed control box, which interfaces with the DCS hardware and with the computer through these motor controller cards.

\fussy
Each motion box and its accompanying linear actuator are powered by and communicate with the computer through a single control box. One control box contains four stepper motor drives\footnote{Kollmorgen. CTDC Stepper Drive (P70530). \mbox{\url{http://www.kollmorgen.com}}} and four 300-W power supplies, one for each of the three motors on the corresponding motion box and one for the linear actuator that controls the motion of the accompanying thermalizer. It also contains a National Instruments Universal Motion Interface (UMI)\footnote{National Instruments. 4-Axis Universal Motion Interfaces for Industrial Applications (UMI-7774). \mbox{\url{http://www.ni.com}}}, which serves as a bridge between the motor controller in the PXI chassis and the hardware in the control box and in the cryostat. The UMI connects to the stepper motor drives, the three motor encoders on each motion box, and the potentiometer that provides position feedback on the linear actuator. It also passes to the PXI card the two micro switch signals from each of the motors in the motion box, the amplified load cell signals that reflect the string tensions, and the state of the thermalizer contact pads (grounded or ungrounded).

All electronic signals for the DCS from inside the cryostat pass through a single wiring port on the cryostat. There are signals from four Cernox thermometers\footnote{Lake Shore Cryogenics. Cernox thin-film resistance cryogenic temperature sensors (CX-1010-SD). \mbox{\url{http://www.lakeshore.com}}. Cernox is a trademark of Lake Shore Cryotronics, Inc.} that measure the temperatures of the sliding blocks of the thermalizers, signals from 12 Cernox thermometers that measure the temperatures of the 12 guide tubes mounted to the top of the \mbox{600-mK} cryostat plate, and signals indicating the four thermalizer contact pad states.

The control boxes each have several output cables that run directly to their respective motion boxes on the cryostat. The linear actuator power and feedback potentiometer signals from all four control boxes, however, are consolidated into only two cables because the linear actuators are all located together near the center of the cryostat. This consolidation occurs inside a small rack breakout box. Also inside the rack breakout box, the thermalizer thermometer signals are directed to a temperature monitor, and the four thermalizer contact pad states are broken out to the four control boxes.

\subsection{Software}
\label{sec:control_software}
Custom software written with LabVIEW\footnote{LabVIEW is a trademark of National Instruments.} controls the entire DCS. With this software, we can operate the system remotely and perform fully automated deployments of the calibration sources into the cryostat.

On the front end, the software displays the complete system status in one of two modes. In the ``visual overview'' mode, the location and direction of motion of the source capsules in the cryostat are shown on a schematic diagram, along with indicators that show when each proximity sensor is recording a capsule and when each thermalizer is in a closed position. In the ``details'' mode, all parameters of the system that are measured and recorded by the software are presented.

The software receives input from the user as a series of text-based commands (``steps''). These steps are collected into procedures, through which the software sequentially progresses when it is in operation. The software is capable of executing multiple procedures in parallel and can synchronize procedures at various points to coordinate the deployments of multiple strings. Thus, we can perform fully automated deployments of the 12 source strings into the cryostat with predefined procedures running in parallel.

On the back end, the software contains a series of interlocks to ensure that the system is operating safely and to prevent the user from accidentally causing any harm to the hardware. Before a string begins to move, the software verifies that the gate valve between the motion box and the cryostat is open and that the thermalizer is also open. If the direction of the requested movement is upward, it checks that the two micro switches in the motion box (i.e. the one that indicates that the string is already in its home position and the one that indicates that the string tension is abnormally high) are not triggered. Before closing a thermalizer, the software also verifies that all three strings that pass through that thermalizer are not moving, and before closing a gate valve, it verifies that all three strings of the corresponding motion box are in their home positions.

The software continuously reads the digital and analog signals from the DCS hardware to ensure that the system is operating as expected. The primary indication that a source string is being lowered or raised through the cryostat correctly is the string tension measured by the load cell. The load cell readings are continuously compared to the load cell profile, which represents the expected load cell reading as a function of the position of the string in the cryostat during a normal deployment. Because each string path through the cryostat is different, there is a unique load cell profile for each string. If the load cell value deviates from a predefined range around this profile for more than 10 seconds, then the string is stopped. The system operator can restart the string motion after assessing the situation, either continuing with or aborting the deployment.

\begin{figure}
\begin{center}\includegraphics[width=3.45in]{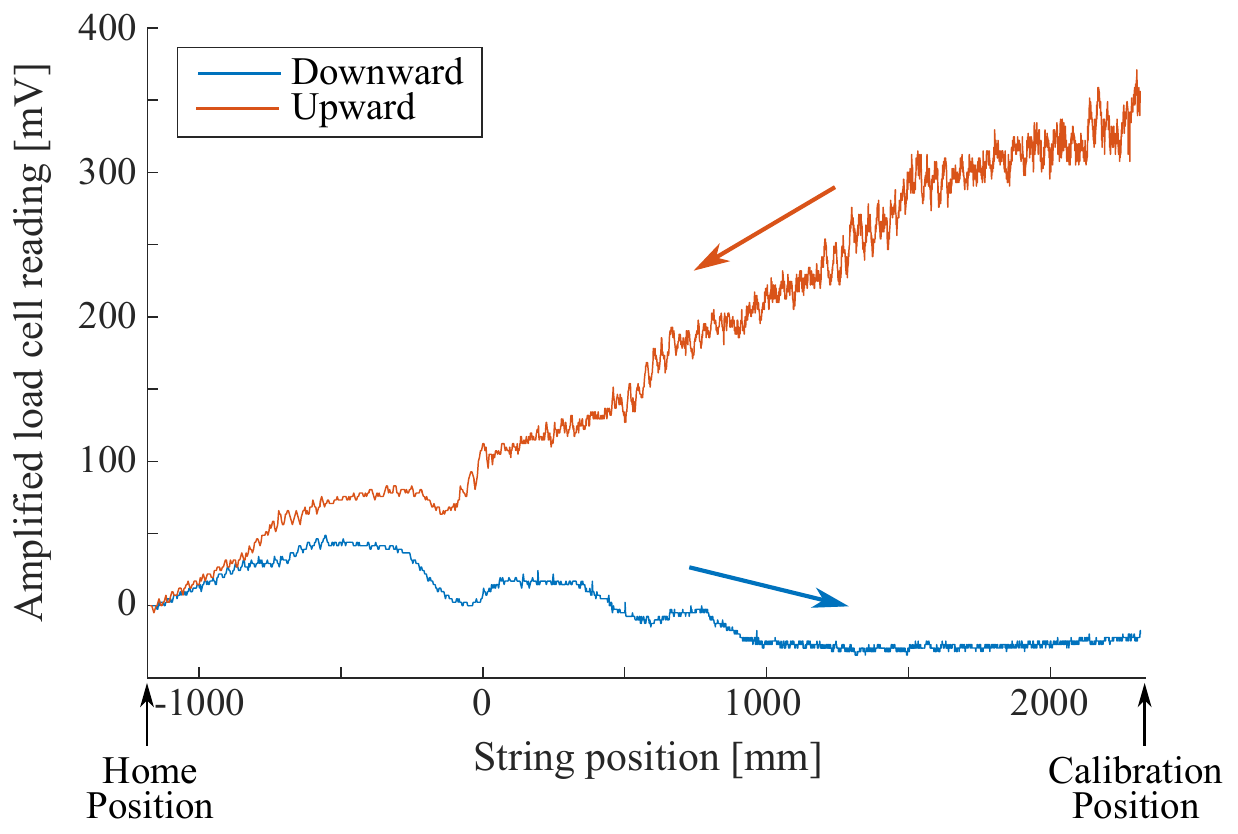}\end{center}
\caption{Load cell readings for a single string during deployment (downward motion) and extraction (upward motion). Downward motion, indicated with the blue arrow, is in the positive direction (left-to-right on this plot). Upward motion, indicated with the red arrow, is in the negative direction (right-to-left on this plot). The tension is much greater in the upward direction because of friction.}
\label{fig:load_cell_deployment}
\end{figure}

A representative load cell plot recorded during the deployment of a single string is shown in \mbox{\autoref{fig:load_cell_deployment}}. As the string is first lowered, the tension is primarily determined by the amount of string that is hanging freely and thus increases approximately linearly. Near the position of $-500$~mm, the bottom of the string encounters the sloped portion of the S~tube, which supports some of the string weight through friction. Once the string reaches higher position values (lower in the cryostat), all capsules have passed completely through the sloping guide tubes and the weight-supporting friction provided by the Kevlar string sliding on the guide tubes is approximately constant. When the string is being withdrawn and is moving in the reverse direction, the tension begins at a much higher value because of the change in the direction of the friction. As the string spools up, progressively less string  is deployed in the cryostat, decreasing the total string mass and total friction of the string in the guide tubes, and the string tension steadily decreases. The tensions for both the upward and downward directions are similar near the home position because once the string is withdrawn from the S~tubes, it is hanging freely and there is little friction. The fluctuations in string tension as the string moves in both directions are repeatable and result from the complex path that each string takes through the cryostat. In particular, the small periodic structures observed during upward motion result from the source capsules moving across bends and other structures as they are withdrawn.

The load cell readings are significantly steadier when the strings are being lowered compared with when they are being raised (see \mbox{\autoref{fig:load_cell_deployment}}). A very tight band of acceptable load cell values is necessary during string lowering to ensure that the string is moving correctly; the string is not actuated from both ends, so the load cell is our primary indicator of successful motion inside the cryostat. We use a band of $\pm 10$~mV around the expected position from the load cell profile, whereas the profile itself varies over $\sim$80~mV during motion in the downward direction. In the upward direction, the band can be much wider, as the only possible exceptional occurrence during a string withdrawal is that the string could become caught or tangled, in which case the tension would significantly increase very rapidly.

Another indicator that the source capsules are successfully entering the guide tubes leading into the cryostat is the proximity sensor at the bottom of each motion box. The software counts the source capsules as they pass through the proximity sensors and into the cryostat, and it resets the corresponding string position value to 0 as the last capsule enters its proximity sensor. It also counts the capsules as they are withdrawn from the cryostat to ensure that all capsules are safely removed from the cryostat after the calibration has concluded.

Finally, the software records the temperatures of the sliding blocks of the thermalizers and the temperatures of the guide tubes anchored to the \mbox{600-mK} cryostat plate. The thermometer on the sliding block of a thermalizer shows a characteristic spike when it squeezes on a warm capsule, which demonstrates that the thermalizer is working correctly. The thermometers on the guide tubes reveal how much heat was not removed by the thermalizer through their temperature rises, and these signals allow us to verify that the sources have actually entered the tubes.

\section{DCS performance in the CUORE Cryostat}
\label{sec:dcs_performance}
We have fully deployed all 12 source strings into the CUORE cryostat in cryostat commissioning runs. The cryostat base temperature during the source deployment tests was approximately 7~mK. During normal operation, the cryostat will be stabilized near 10~mK, a temperature that was also used for \mbox{CUORE-0}~\cite{Alfonso:2015vk}. Our testing was designed to find the maximum possible deployment and extraction speeds without causing the base temperature to rise above 9~mK, which, in these tests, corresponds to a heat load on the mixing chamber of approximately \mbox{2--3~$\mu$W}~\cite{Chott:2014id}. This provides the temperature stabilization system with sufficient leeway to maintain a constant base temperature of 10~mK during string deployment, leaving the detectors undisturbed.

\subsection{Deployment}

Cooling the sources down from room temperature to 4~K is a slow process. To this end, we left the source capsules in the S~tubes for a full day before lowering them further into the cryostat, a process we refer to as ``precooling.'' The S~tubes maintain a thermal gradient from 300~K to 4~K, and precooling of the strings is therefore achieved through contact with the sloped walls of the tubes. Simulations indicate that the background contribution in the $0\nu\beta\beta$ region of interest from having the sources in the cryostat but fully above the thermalizers is under $0.2\%$ of the CUORE background budget. Thus, normal low-background data taking can continue during precooling.

The deployment procedures for the inner and outer source strings begin in a similar manner. Following precooling, the \mbox{4-K} thermalizer squeezes pairs of adjacent capsules on the strings, each for a period of time ranging from 10 to 20~minutes. The string is lowered by 58~mm (twice the pitch of the capsules on the string) between each squeeze. Longer squeezes are required for the higher capsules, which are precooled to higher temperatures because of their relative positions along the thermal gradient in the S~tubes. In total, the thermalizer squeezes last for approximately 4~hours per string.

\begin{figure*}
\begin{center}\includegraphics[width=6in]{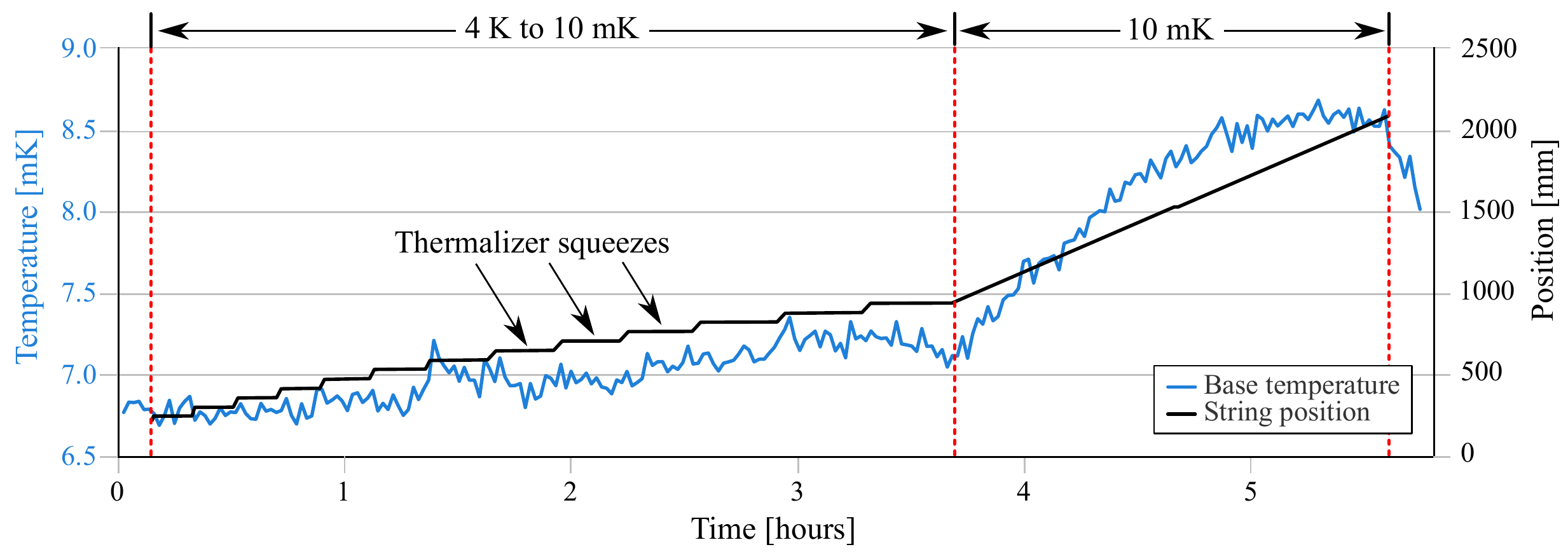}\end{center}
\caption{Cryostat base temperature and string position during a full inner-string deployment. Downward motion is in the positive direction. Two regions are identified; the first consists of progressively longer thermalizer squeezes while the bottom of the string moves from the \mbox{4-K} cryostat plate to the \mbox{10-mK} plate, and the second is the movement of the string through the \mbox{10-mK} plate to its final deployment position.}
\label{fig:inner_deployment}
\end{figure*}

For the inner strings, following the final thermalizer squeeze, the bottom of the source string is approximately at the level of the \mbox{10-mK} cryostat plate. Before the strings reach the coldest stage of the cryostat (the mixing chamber level), they have very little effect on the cryostat base temperature. Indeed, a single inner string causes the cryostat to warm up only from 6.7~mK to 7.2~mK during this phase of deployment (see \mbox{\autoref{fig:inner_deployment}}). We are able to perform this thermalization sequence on four strings simultaneously while keeping the cryostat temperature below 9~mK. The outer strings have little to no effect on the cryostat base temperature during this phase of deployment.

The second deployment phase for each string involves lowering it from its final squeeze position to its full deployment position. It is in this phase that the inner strings are at or below the level of the mixing chamber on the \mbox{10-mK} plate. We must bring the strings into this region very slowly to avoid exceeding the cooling power of the dilution refrigerator and causing the base temperature to spike. In testing, we determined that a single inner string moving at 10~mm/minute into this region raised the base temperature to 8.6~mK (see \mbox{\autoref{fig:inner_deployment}}).

The outer strings are deployed outside of the \mbox{50-mK} vessel and thus do not impact the base temperature directly. The important parameter for the outer strings is the \mbox{50-mK} vessel temperature (and, by proxy, the temperature of the heat exchanger of the dilution refrigerator), which begins to affect the base temperature if it rises and remains above \mbox{$\sim$80--100~mK}. We determined that up to four outer strings can be simultaneously deployed at speeds of 15~mm/minute with only minor effects on the cryostat base temperature.

The most sensitive and time-consuming part of the deployment is the second deployment phase of the inner strings, which must be done serially. Following a parallelized four-hour first deployment phase, this second phase requires approximately two hours for each inner string, corresponding to a total of approximately 12 hours for all six inner strings. During the entire process, whenever a motion box is not moving an inner string, we can deploy one of its outer strings. The motion boxes are capable of moving all 12 strings simultaneously, but the \mbox{4-K} thermalizers squeeze on all three string paths below a motion box simultaneously, thus making it impossible to move one string while squeezing on another from the same motion box. With a properly planned strategy, this is not a limiting factor in our deployment time.

\subsection{Extraction}

Following the calibration period, we extract the source strings from the cryostat. Because the sources have had time to equilibrate with the cryostat and fully cool, the heat load on the cryostat from the string extraction originates almost entirely from friction. Although there is no need to squeeze on the source capsules with the \mbox{4-K} thermalizers during string extraction, the slow speeds required to avoid excessive frictional heating limit the total extraction time to only slightly less than the deployment time.

\begin{figure*}
\begin{center}\includegraphics[width=6.5in]{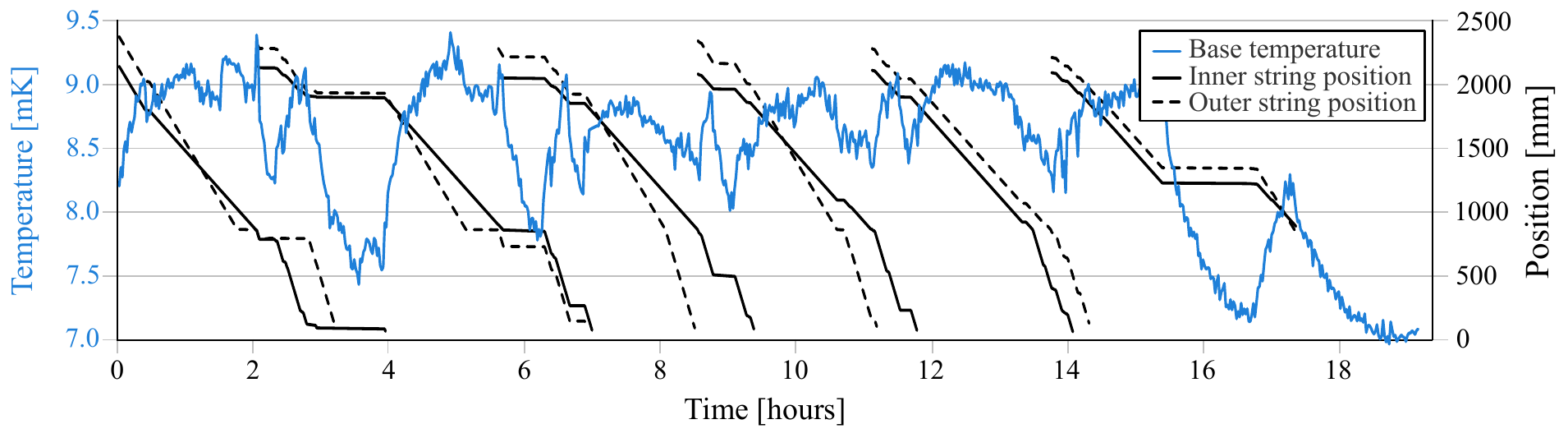}\end{center}
\caption{Cryostat base temperature and string positions during a 12-string extraction. The 12 downward-going black lines represent the positions of the 12 strings (upward motion is in the negative direction). In this procedure, the strings were extracted from their deployment positions to their precooling positions near the 4-K plate. Horizontal segments in the lines representing the string positions correspond to periods of no motion to observe the temperature response and recovery time of the cryostat.}
\label{fig:string_extraction}
\end{figure*}

As during string deployment, the most time-consuming part of the string extraction is the movement of the inner strings in the region below the \mbox{10-mK} plate. A single inner string extracted at 10~mm/minute, in parallel with an outer string extracted at 15~mm/minute, raised the base temperature to approximately 9~mK during the time when it was partially below the mixing chamber plate. Thus, we extract the strings in pairs, each consisting of one inner and one outer string, and can begin to extract the next pair of strings when the previous pair is above the mixing chamber. In the warmer parts of the cryostat, extraction can continue at significantly greater speeds without impacting the base temperature. A 12-string, 16-hour extraction is represented in \mbox{\autoref{fig:string_extraction}}.

\section{Conclusion}

We have designed and constructed the CUORE Detector Calibration System to perform periodic {\it in situ} energy calibrations of the CUORE detector towers. The DCS meets the thermal and radioactivity requirements of the CUORE cryostat while enabling controlled, reliable calibration source deployment. We have also developed an electronic control system connected to a dedicated server to operate the DCS and perform fully automated calibration sequences. During calibration periods, sensors monitor the DCS to ensure the safety of the calibration sources and the cryostat.

We performed a full system test of the DCS during the CUORE cryostat commissioning period. The DCS can deploy all source strings while allowing the cryostat to maintain a base temperature below 10~mK, and it can extract them following each calibration period with similar effects. Thus, we have verified that the DCS has the ability to perform a full calibration sequence without affecting the temperature and operation of the detectors in the temperature-stabilized cryostat.

\section{Acknowledgments}
We would like to thank Ken Kriesel and the University of Wisconsin-Madison Physical Sciences Laboratory for their assistance with the design and machining of the motion boxes and other hardware for the DCS. We would also like to thank Larry Bartoszek, Jess Clark, Alberto Franceschi, Ian Guinn, Tommaso Napolitano, Marco Olcese, Basil Smitham, Lauren Wielgus, and Adam Woodcraft for their contributions to the early design and R\&D work on the system, as well as Keenan Thomas for measuring the activity of the source material. In addition, we thank Carlo Bucci, Suryabrata Dutta, Paolo Gorla, Ke Han, Tommy O'Donnell, Vivek Singh, Joe Wallig, and many other members of the CUORE Collaboration for their valuable assistance in the commissioning and testing of the DCS at the Laboratori Nazionali del Gran Sasso.

This work was supported by the Istituto Nazionale di Fisica Nucleare, the Alfred P. Sloan Foundation, the University of Wisconsin Foundation, and Yale University. This material is based upon work supported by the US Department of Energy, Office of Science, Office of Nuclear Physics, under Award Number DE-SC-0012654.

\bibliography{DCSpaper}

\end{document}